\documentstyle[12pt,aaspp,tighten]{article}

\def\fun#1#2{\lower3.6pt\vbox{\baselineskip0pt\lineskip.9pt
  \ialign{$\mathsurround=0pt#1\hfil##\hfil$\crcr#2\crcr\sim\crcr}}}

\def\VEV#1{\left\langle #1\right\rangle}

\def\mass{{\cal M}}

\def\Msolar{{\mass_\odot}}
\def\Ampl{{\cal A}}
\def\thresh{{\rm thresh}}
\def\max{{\rm max}}

\slugcomment{\vbox{\hbox{CU-TP-715}
             \hbox{CAL-583}}}

\begin{document}
\title{Rates for Color Shifted Microlensing Events}

\author{Ari Buchalter$^*$\altaffilmark{1}, Marc
Kamionkowski$^\dagger$\altaffilmark{2}, and R. Michael
Rich$^*$\altaffilmark{3}}
\affil{$^*$Department of Astronomy, Columbia University, New
York, NY 10027}
\affil{$^\dagger$Department of Physics, Columbia University, New
York, NY 10027}
\altaffiltext{1}{ari@parsifal.phys.columbia.edu}
\altaffiltext{2}{kamion@phys.columbia.edu}
\altaffiltext{3}{rmr@cuphya.phys.columbia.edu}

\begin{abstract}

If the objects responsible for gravitational microlensing of
Galactic-bulge stars are faint dwarfs, then blended light from
the lens will distort the shape of the microlensing light curve
and shift the color of the observed star during the microlensing
event.  In most cases, the resolution in current
microlensing surveys is not accurate enough to observe this color-shift
effect.  However, such signatures could conceivably be
detected with frequent followup observations of microlensing events in
progress,
providing the photometric errors are small enough. We calculate the
expected rates for microlensing events where the shape
distortions will be observable by such followup observations,
assuming that the lenses are ordinary
low-mass main-sequence stars in a rapidly rotating bar and in
the disk.  We adopt Galactic models consistent with observed microlensing
timescale distributions, and consider separately the cases of self-lensing
of the bulge, lensing of the bulge by the disk, and self-lensing
of the disk, further differentiating between events where the source is a giant
or a main-sequence star. We study the dependence of the rates for
color-shifted microlensing events on the
frequency of followup observations and on the precision of the
photometry for a variety of waveband pairings.
We find that for hourly observations in $B$ and $K$
with typical photometric errors of 0.01 mag, 28\% of the events where a
main-sequence bulge star
is lensed, and 7\% of the events where the source is a bulge giant, will give
rise to a
measurable color shift at the 95\% confidence level. For observations in $V$
and $I$, the fractions become 18\% and 5\%, respectively, but may be increased
to 40\% and 13\% by improved photometric accuracy and increased sampling
frequency.
Unlike standard achromatic lensing events,
color-shifted events provide information on the lens mass, distance, and
velocity. We outline how these parameters can be obtained, giving examples of
typical
errors which may arise in the calculation, and briefly discuss other
applications
of such light-curve measurements. We show that color-shifted events
can be individually and/or statistically distinguished from events where the
source is blended with a binary companion.
In particular, the observed fraction of color-shifted events as a function of
event timescale
will test whether the color shift is caused by a lens which is a low-mass
main-sequence star,
or by a blended star.

\end{abstract}



\section{INTRODUCTION}

The MACHO (Alcock et al.~1993, Alcock et al.~1995a), EROS (Aubourg et
al.~1993),
OGLE (Udalski et al.~1994a), and DUO (Alard et al.~1995)
gravitational-microlensing surveys
have yielded a wealth of interesting but puzzling results.
Although the rate of microlensing of stars in the LMC appears to
be too small for the lenses to  account for the dark matter in
the Galactic halo needed to support the Galactic rotation curve, the
nature and location of the 6 MACHO lenses observed to date toward the
LMC remain a mystery.  Are they in a ``maximal disk,'' the halo, or
perhaps in the LMC itself? Even more intriguing are the results
of surveys of the Galactic bulge: the MACHO and OGLE
groups have to date observed 85 and 12 lensing events, respectively,
toward the bulge. The optical depth to
microlensing toward the bulge appears to be 2-3 times higher
than expected from lensing by low-mass stars in simple Galactic models.
One possibility is that this enhancement is due to
a hitherto undiscovered population of compact dark objects which
suggests the existence of more mass in the disk than previously
believed (Alcock et al.~1994, Gould 1994a).  Another explanation is that
the larger optical depth
observed is due to structure, such as a bar, in the bulge
(Kiraga \& Paczy\'nski 1994, hereafter KP; Zhao, Spergel, \& Rich 1995,
hereafter
ZSR).  Several lines of evidence seem to point to the latter
possibility (Binney et al.~1991; Blitz \& Spergel 1991; Whitelock \& Catchpole
1992;
Stanek et al.~1994; Dwek et al.~1995). In
this case, the lenses would mostly be low-mass, but otherwise
ordinary, stars in the bulge itself. Thus, gravitational microlensing is
proving to
be a primary tool in the understanding of Galactic structure
and in the study of stellar populations in the bulge and inner Galactic disk.

The optical depth to microlensing,
the fraction of source stars undergoing
microlensing at any given time, can be measured from the observed
timescales (i.e., the event durations) and the frequency
with which events occur, to give an approximation
to the integrated mass in lenses along the line of sight.
However, the timescale
\begin{equation}
     t_0={R_e \over v}; \qquad R_e=\left( {4 G \mass_{l}
     D_{ol} D_{ls} \over c^2 D_{os}} \right)^{1/2},
\label{timescaleequation}
\end{equation}
(the time the source star remains within the Einstein
ring of the lens), where $R_e$ is the Einstein
radius, depends on the
distances from the observer to both the source star and the
lens, $D_{os}$ and $D_{ol}$, the
transverse speed $v$ of the lens relative to the
source-star line of sight, and the mass $\mass_{d}$ of the
lens (and $D_{ls} \equiv D_{os}-D_{ol}$).  Although the
distances to the source stars in the LMC, and to a
lesser extent in the bulge, are known to reasonable accuracy, the
mass, speed, and distance of any given lens cannot be
disentangled in any given event.

Several techniques for breaking the degeneracy in $t_0$ have
been proposed.  Measurement of the amplification of limb
brightening in giant source stars during a microlensing event
could determine $R_e/D_{ol}$ and $v$ (Loeb \& Sasselov 1995).
The combination, $v/D_{ol}$, can be measured photometrically
(Gould 1994b; Nemiroff \& Wickramasinghe 1994) and
spectroscopically (Maoz \& Gould 1994).  Measurement of
microlensing parallaxes from a single satellite in heliocentric
orbit could determine the ``reduced'' transverse speed and
Einstein radius, $(D_{os}/D_{ds})v$ and $(D_{os}/D_{ds}) R_e$,
respectively (Gould 1995a; Han \& Gould 1995a).
For the rare events with timescales comparable to a year,
measurement of parallax from the ground can already be used to
measure the reduced velocity, and such parallax shifts may be observable
in a larger fraction of events with frequent and
accurate followup observations (Buchalter \& Kamionkowski 1995, in
preparation).
All of these techniques provide
more information than that provided by just the timescale,
although they still leave a two-fold degeneracy.  (The
degeneracy could be broken completely by a second parallax
satellite.)  Realistically, however, some of these ideas are
effective only in a small fraction of events, and others
will require several years to come to fruition.

Recent calculations of microlensing towards the bulge indicate
that Galactic models having a mass function with 25\% or
more of the mass in the form of brown dwarfs are inconsistent
with the current MACHO and OGLE data; the best-fit models have a
number-averaged mean mass of 0.3 $\Msolar$ (ZSR; Zhao, Rich, \& Spergel 1995,
hereafter
ZRS).
If the lenses are indeed low-mass main-sequence stars,
then another possibility exists for removing the
degeneracy (Kamionkowski 1995).  In that case,
both the source star and lens fall within a single
seeing disk and the light from the two
becomes blended.  However, during the microlensing event, only
the light from the source star is amplified.  Stars generally
have different colors, so during the event, the color of the two
blended stars should change.  Furthermore, the contribution of
unlensed light from the lens will distort the shape of the
standard microlensing light curve, even if there is no color
difference.  Accurate measurement of
the light curves of
color-shifted events (CSEs) can be used to obtain the brightness of the
lens in two or more wavebands in which case the mass,
distance, and transverse speed of the lens could be
determined through spectroscopic parallax.

This color shift is unlikely to be dramatic
enough to be observed by any of the existing surveys in all but
a small fraction of events; the frequency with which the light
curves are currently sampled is too small, and the photometric
errors are too large.  For existing data, color-shift analysis can only be used
to constrain
the lens mass as a function of distance (Kamionkowski 1995; Kamionkowski \&
Buchalter 1995; Alcock et al.~1995b).
However, both the OGLE and MACHO
collaborations have now developed an early-warning alert system
which can signal a potential microlensing event in progress shortly after
the light curve begins to rise (Stubbs et al.~1994; Udalski et al.~1994b).
Therefore, it is conceivable
that a program of followup observations of microlensing light
curves in progress with frequent (e.g., hourly) measurements
and small photometric errors could be used to
measure the light curve with sufficient accuracy to pick out
color shifts and shape distortions in a significant fraction of
the events.  In fact, such a followup program is already being
considered with the primary purpose of detecting planets around
the lenses (Tytler et al.~1996, in preparation). Planetary masses
would give rise to smaller typical timescales (on the order of 2 to
50 hours) and thus produce narrow spikes on the lensing light curve which
likely go undetected in current observations. These spikes could be resolved by
dedicated telescopes performing rapid sampling of events in progress, with
low photometric errors. The same survey would thus be
ideal for the detection of color-shifted distortions.

In this paper, we evaluate the fraction of microlensing events
toward the Galactic bulge which should have an observable shape
distortion in such a followup program.  We calculate the
fractions that will arise if the lenses are all in the bulge and
if the lenses are all in the disk. We consider surveys which
focus only on source stars which are giants, and those in which
the source stars are main-sequence stars. To do so, we use several realistic
models for the distribution
of lenses, with a timescale distribution which matches
roughly that for the events observed so far.  A Monte Carlo technique is
then used to simulate events and generate shape-distorted
microlensing light curves for each.  It is then determined for
each event whether the shape-distorted light curve can be
distinguished at the 95\% confidence level from a standard
achromatic microlensing light curve.  The calculation is
performed for several viable values of the sampling frequency
and for several values for photometric accuracies that are
realistically attainable.  The final results are displayed as a
function of sampling frequency for several different levels of photometric
error.

Our results indicate that dedicated telescopes sampling
events in progress at an hourly rate in $B$ and $K$ should observe a color
shifplanetaryt for
roughly 30\% of the events where a main-sequence bulge star is
lensed (by either a disk or bulge star) given photometric errors of
0.01 mag. For the lensing of a bulge giant,
the expected fraction drops to 7\%. For rarer events where a disk
source is lensed by a foreground disk star, the calculated rate of color
shifting
is 51\% for dwarf sources and 8\% for giants. We find
comparable results for observations in
$V$ and $I$, yielding values typically 0.6 to 0.7 times those for $B$ and $K$.
The expected fraction of CSEs increases significantly as the photometric errors
and
sampling intervals are reduced; for data sampled every fifteen minutes in $V$
and
$I$ with 0.005 mag errors, 40\% of the events with bulge main-sequence sources
and 13\% of those with bulge clump-giant sources are expected to be color
shifted.
Events with this characteristic distortion to the microlensing light curve,
including those where a  mass is discovered, can be studied to directly
infer the locations, masses, and transverse speeds of
the lenses to within relatively small errors.
Furthermore, light curves distorted by a luminous lens may be distinguished by
various
techniques from those where the source is merely blended with another star
along the line of sight.
We emphasize that the color-shifting effect must be observed by the intensive
monitoring surveys if the lenses are low-mass dwarfs, and would thus provide
direct
evidence that normal stars are responsible for microlensing towards the bulge.
In addition, color-shift analysis can be applied immediately to all events,
so that information about the mass, velocity, and spatial distributions of the
lensing population will
become instantly accessible in many cases, without
the need for long-term followup observations.

The plan of the paper is as follows:  In Section 2, we review
the characteristics of the shape distortions to the light curves.
In Section 3, we outline the model used for our calculation,
including the bulge and disk models employed, and the
relations used to connect brightnesses
with stellar masses.
In Section 4, we calculate the
timescale distribution for these models and compare with the
currently observed distributions.  Section 5 discusses the
criteria used to determine whether a given event can be
distinguished from a standard microlensing light curve with
sufficient statistical significance.  In Section 6, we
consider what may be learned from existing data and present
the results of our calculations of the expected fractions of CSEs.
Section 7 addresses how well light-curve analysis in such experiments will be
able to
constrain the lens mass, distance, and speed. In Section 8, we discuss
methods of distinguishing color-shifted events from ordinary blended events,
and in Section 9 we make some concluding remarks.

\vfill\eject

\section{COLOR-SHIFTED MICROLENSING EVENTS}

Consider first ordinary achromatic microlensing by a nonluminous
lens.
If the distance of the lens from the source-star line of
sight is $R_e u$, where
\begin{equation}
     R_e = \left[ {4G\mass_l D_{ol} (D_{os} - D_{ol}) \over c^2 D_{os}}
     \right]^{1/2}
      = 3.2 \, {\rm a.u.}\, \left[ {\mass_l \over \Msolar} {D_{os}
     \over 8\, {\rm kpc}} { x' \over 0.8} {1-x' \over 0.2}
     \right]^{1/2}
\label{einsteinradius}
\end{equation}
is the Einstein radius of the lens and $x'=D_{ol}/D_{os}$, then the
amplification of the source star as a function of time $t$
is
\begin{equation}
     A[u(t)] = \frac {u^2 +2} {u(u^2 +4)^{1/2}};\qquad  u(t)=
     \left[  \left( \frac {t-t_{\max}} {t_0} \right)^2 +
     u_{\min}^2\right]^{1/2},
\label{standardamplification}
\end{equation}
where $u$ is the impact parameter in units of the Einstein
radius, ${t_0} = R_{e}/v$ is the event timescale, $v$ is the
transverse speed of the lens relative to the source-star line of
sight, and $t_{\max}$ is the time at which peak amplification,
$A_\max = A(u_{\min})$, occurs.

A microlensing event is detected when the amplification
exceeds $A_T$ which corresponds to a
dimensionless lens--line-of-sight distance of $u_T$ (e.g.,
$A_T=1.34$ for $u_T=1$).  A standard achromatic event is
described by three parameters: $u_{\min}$ (or equivalently,
$A_\max$), the timescale $t_0$,
and the time $t_{\max}$.
The event duration $t_{e}$ is the time that
the amplification is above threshold, and it is also the time
the lens remains within a distance $u_T R_e$ from the line of
sight; it is related to the timescale ${t_0}$.
The system
parameters (the distances to and masses of both stars and the
transverse speed) cannot be determined uniquely in any
given event.  These quantities can in principle be disentangled
in a statistical manner if a number of events are observed, and
only with some assumptions about the mass, spatial, and speed
distributions of the lenses.
Microlensing is a gravitational effect, so the amplification is
the same in all wavelengths.  Therefore, if the lens is too
faint to be observed, the event will be achromatic.

If the lens is itself an ordinary (although perhaps
low-mass) main-sequence star, then it will emit light of its
own.  Therefore, let us now consider the effect of unlensed
light from the lens on the microlensing light curve.
The brightness of a star of luminosity $L$ at a
distance $d$ is $\ell = L/d^2$.  The source star with brightness
$\ell_s$ is lensed by a star with brightness $\ell_l$.
The light observed is a
combination of the light from both stars, and the colors of the
two stars will generally be different (Griest \& Hu 1992; Gould
\& Loeb 1992; Udalski et al.~1994c; Kamionkowski 1995).
When the lens passes within the microlensing tube, the
source star will be
amplified, and the relative brightnesses will change.
Consequently, the color will change, and achromaticity is lost.  To
properly describe the event, we must consider the brightness
in each waveband separately.  If the amplification of the
background star is $A$, then the brightness in waveband
$\lambda$ is $\ell_{\lambda} = \ell_{l\lambda} + A
\ell_{s\lambda}$,
where $\ell_{l\lambda}=L_{l\lambda}/d_l^2$ is the brightness
of the lens star in $\lambda$, $L_{l\lambda}$ is the
luminosity of the star in $\lambda$, and similarly for the
source star.  The baseline brightnesses are obtained by
setting $A=1$.

Although $A$ is the amplification of the
source star, the observed light in this
case comes from both stars, so the {\it observed} amplification,
which we denote by $\Ampl$, is different from the microlensing
amplification $A$.  Although the amplification of the background
star is indeed achromatic, the observed amplification will
depend on wavelength, and in waveband $\lambda$ it is
\begin{equation}
     \Ampl_\lambda (t)= {\ell_{l\lambda} + A(t) \ell_{s\lambda}
                    \over \ell_{l\lambda} + \ell_{s\lambda}}
                = (1-r_\lambda) + A(t) r_\lambda,
\label{observedamplification}
\end{equation}
where $r_\lambda=\ell_{s\lambda} / (\ell_{s\lambda} +
\ell_{l\lambda})$ is the luminosity offset ratio (note $r_\lambda=1$
for a completely dark lens).
Suppose, for example, light curves are measured in two bands,
$\alpha$ and $\beta$.  If $\ell_l \ga \ell_s$ (in both
bands), then the baseline color observed will be
$\ell_{l\alpha}/\ell_{l\beta}$ (the color of the
lens), but in a lensing event with amplification $A\gg1$, the
observed color will be $\ell_{s\alpha} / \ell_{s\beta}$ (the
color of the background star). In general, a color-shifted event will be
registered when
the {\it observed} amplification is greater than threshold,
$\Ampl > A_T$.  This depends on the band,
so strictly speaking, the exact time at which an event is
registered (or whether it is registered at all) may depend on
the waveband.  If events are triggered by the light curve in
$\lambda$, they will be registered when the microlensing
amplification is $A \geq A_{\rm thresh}=[A_T - (1 - r_\lambda)]/
r_\lambda$, or only when the dimensionless lens--line-of-sight
distance is $u \leq u_{\rm thresh} = u(A_\thresh)$.
The duration of an event, $t_e = 2{t_0}[ u_\thresh^2 -
u_{\rm min}^2 ]^{1/2}$, depends on the transverse
speed, the mass of the lens, the distances to both stars
through ${t_0}$, the brightnesses of both stars through the $r$
dependence of $u_\thresh$, and on the impact parameter.  In all cases,
the duration of
such color-shifted events should be shorter than the duration of
achromatic events (Kamionkowski 1995).

Figure 1 illustrates an example of a microlensing light curve distorted by
the color-shift effect. Note that the different light curves have
the {\em same} values of $t_0$ and $u_{\rm min}$, and differ only
in $r_{\lambda}$.
\begin{figure}
\plotone{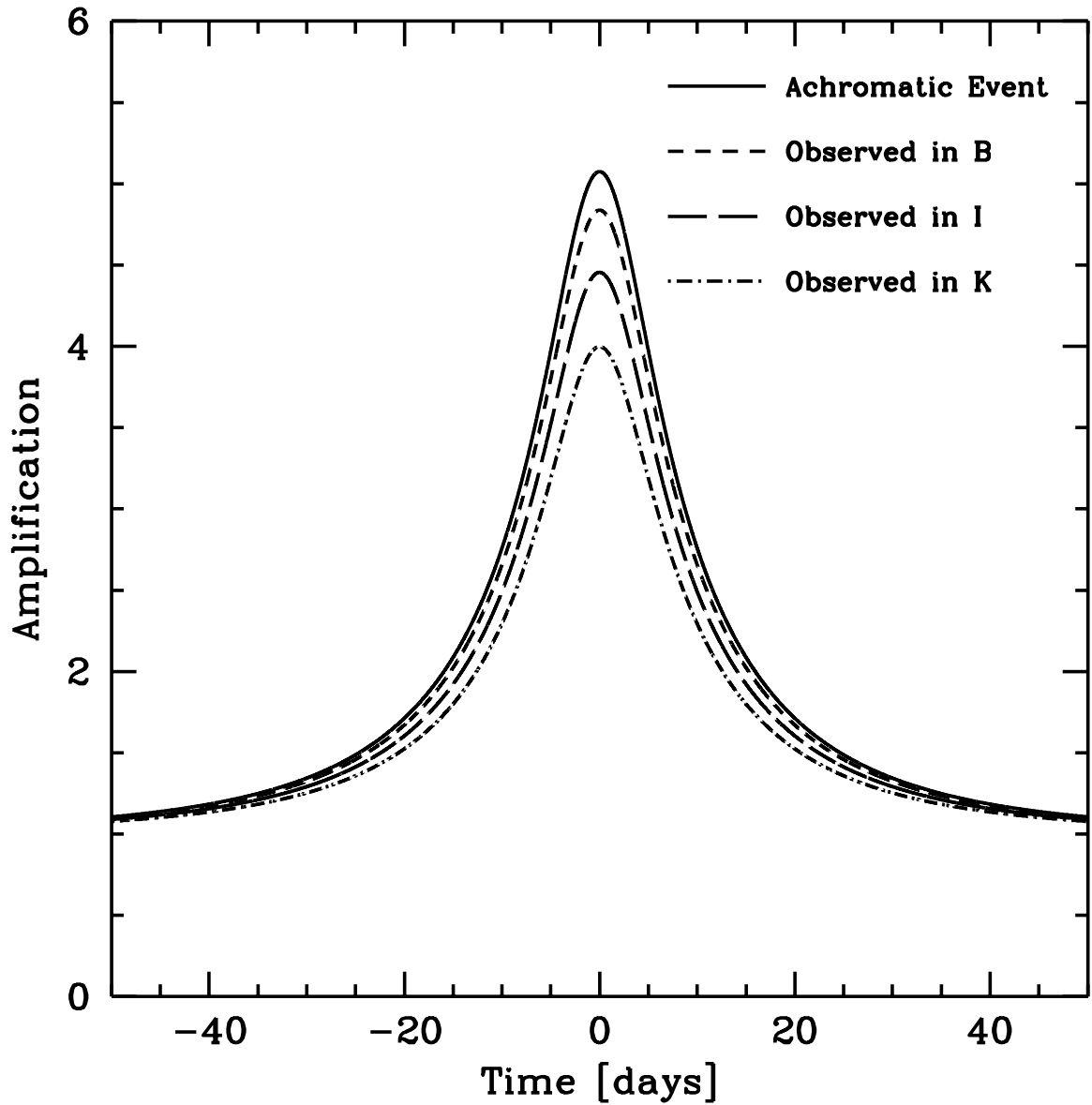}
\caption{Microlensing light curve for
a 0.6$\Msolar$ object at 4 kpc lensing a 1$\Msolar$ main-sequence
star at 8 kpc. The solid curve represents the standard achromatic
amplification (i.e., for a ``dark'' lens) in any waveband. The dashed curves
represent the same event observed in different wavebands if we assume
the lens is a main-sequence star. The light curves were all generated with
the same $t_0$, $t_{\rm max}$, and $u_{\rm min}$, but with differing values
of $r_\lambda$ in the different bands as given by the assumed stellar
mass-luminosity relations (see below).}
\label{ampfig}
\end{figure}
In addition to the shift
in amplification, the contribution of the unlensed light
changes the {\it shape} of
the microlensing light curve in any given band as well. In Figure 2,
we plot a simulated color-shifted light curve with
$t_0 = 30$ days, $u_{\rm min} = 0.1$ and $r_{\lambda} = 0.5$.
This curve has a peak amplification of 5.52 and a FWHM of 12.1 days.
We also plot an achromatic light curve (i.e., only 3 parameters) with the same
peak height and FWHM
(both curves are assumed to have $t_{\rm max} = 0$). The
two curves and the residuals show that they
have different shapes, and
that the effect of color shifting is observable along the
{\it entire} light curve.
\begin{figure}
\plotone{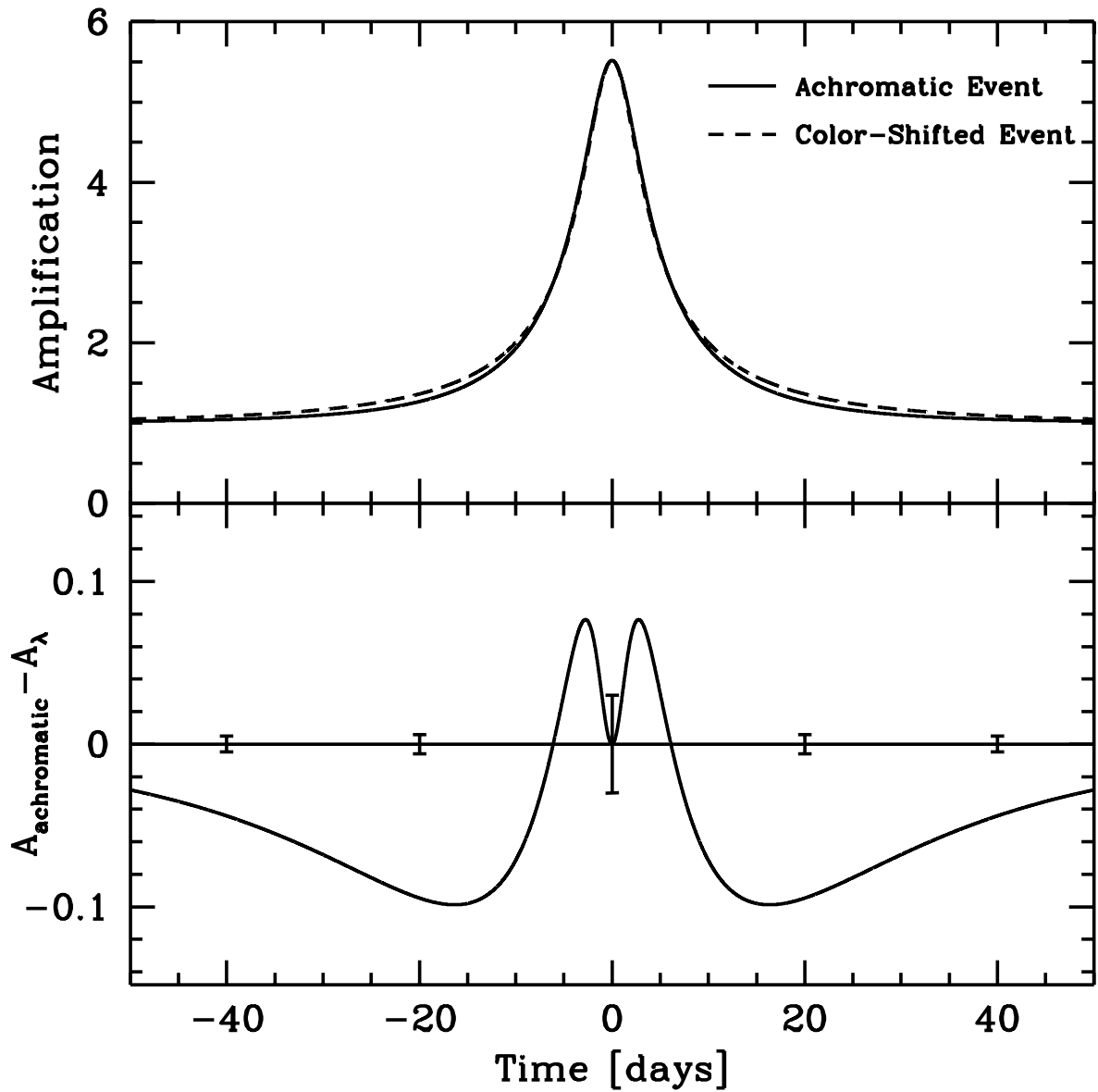}
\caption{The upper panel shows a color-shifted light curve (dashed line) with
parameters $t_0 = 30$ days, $u_{\rm min} = 0.1$, and $r_{\lambda} = 0.5$,
and an achromatic light curve (solid line) with identical peak height and FWHM.
The
achromatic curve has $t_0 = 18.12$ days and $u_{\rm min} = 0.18$. The lower
panel shows the residuals in $A_{\rm achromatic} - A_{\lambda}$, along
with typical expected errors for the proposed planetary surveys.}
\label{residuals}
\end{figure}
It is important to note that
if a color-shifted lensing event is fit for only the three achromatic
parameters $t_0$, $t_{\max}$, and $A_{\max}$, the inferred values can
be substantially different from their true values. In this case the
best achromatic fit to the color-shifted curve yields the incorrect values of
$t_0 = 18.12$ days
and $u_{\rm min} = 0.18$, values which are in error by roughly 40\% and 80\%,
respectively.
The ``true'' achromatic curve corresponding to the event would have
a larger peak height and FWHM, as in Figure 1.

If both the source and lens are main-sequence stars, then
the source is usually brighter than the lens in most of the
events that will be detected (Kamionkowski 1995).  Moreover, if
the sources are giants, the source will almost always be much
brighter than the lens.  As a result, color shifting and
shape distortions to the light curves will usually be
small.  Therefore, we will heretofore neglect the small
effect of unlensed light on the threshold amplification and
event duration.

\section{GALACTIC MODEL}

Gravitational microlensing  has rapidly become an important
tool for understanding Galactic structure. Yet despite the
large event rate observed toward the Galactic bulge, the basic
characteristics of the lens population are unknown due to the
simultaneous dependence of the relevant parameters $\mass_l$, $D_{ol}$,
and $v$ on the fitted parameter $t_0$. Observation of shape distortions
to microlensing light curves may help break the
degeneracy. In particular, such distortions can potentially be observed
using high-precision, intensive monitoring. Therefore, we construct
a model to calculate the expected fraction of events along the
line of sight to Baade's Window (${\ell} = -1^{\circ}, b=4^{\circ}$) in which a
color-shift distortion will arise.
We consider separately three scenarios of self-lensing of the Galactic bulge,
lensing of the bulge by the disk, and self-lensing of the disk. Here we
describe the bulge and disk models, as well as the
mass-luminosity and color-color relations employed in the calculation. The halo
is not expected to make
any significant contribution to the optical depth toward the bulge.

\subsection{Bar Model}

For the bar, we use the model of Zhao (1995), a
numerically-constructed, self-consistent model of a rapidly-rotating
bar.  The density profile matches the Dwek et
al. (1995) fit to the COBE image of the Galaxy, and the
velocity dispersions match those inferred from measured
proper motions (Spaenhauer, Jones, \& Whitford 1992) and radial
velocities (Rich 1988, 1990). (See Zhao 1995 and ZRS for a more
detailed discussion). The bar model consists of 2,308 stars generated from a
Monte Carlo
of Zhao's phase-space distribution,
each with a distance along the line of sight to Baade's Window, and proper
motions $v_{\ell}$ and $v_b$.\footnote{We thank H. S. Zhao for providing this
sample.}

Our analysis of CSEs depends on the stellar mass functions.
Though the luminosity function of the bulge is not well constrained,
several groups have used microlensing observations to fit the bulge mass
function (MF).
Both ZRS and Han \& Gould (1995b), using data from 55 MACHO and OGLE events,
find
the best fitting MF to be a power law ($dN/d\mass \; {\propto} \; \mass^{-p}$)
with an index $p$ between
2 and 2.5, and a lower mass cutoff near 0.1 $\Msolar$. Following KP, we adopt a
logarithmic MF,
$dN/d\mass \; {\propto} \; \mass^{-2}$, in the range $0.1 {\Msolar} \leq
{\mass_l} \leq
1.2 {\Msolar}$, which is consistent with the observed MACHO and OGLE timescale
distributions (see ZRS Figure 2).

\subsection{Disk Model}

For the galactic disk, we adopt a double exponential profile with
\begin{eqnarray}
	\rho_d & {\propto} & \exp\left( - {R}/{h_R} \right)\exp\left( - {z}/{h_z}
\right),
\end{eqnarray}
where $\rho_d$ is the density of stars, $R$ and $z$ are Galactocentric radial
and vertical coordinates and $h_{R}=3.5$ kpc
and $h_z=350$ pc are the radial and vertical scale lengths. We assume the value
$\rho_{d,\odot} = 0.1 \Msolar \mbox{pc}^{-3}$ at a solar Galactocentric
distance
of 8 kpc. The color-magnitude diagram (CMD) of stars in bulge fields shows a
deficit of foreground disk stars (Paczy\'{n}ski et al.~1994). Thus, following
KP and ZRS, we truncate the disk
beyond 5 kpc from the solar circle, leaving a central hole of radius 3 kpc. The
corotation
radius of the bar extends only to 3 kpc from the Galactic center, so the disk
and bulge stars in our model can be distinguished by their distance alone.
We assume a constant disk rotation speed of $v_{\phi} = \mbox{220 km s}^{-1}$
(Merrifield 1992). Lewis \& Freeman
(1989) measured the velocity dispersions of disk stars and found relations of
the
form ${\sigma_{R,\phi}(R)}={\sigma_{R,\phi}}(0)\exp\left[ - {R}/{2h_{R,\phi}}
\right]$
for the radial and azimuthal velocity dispersions, and it is assumed that
${\sigma_z}(R) \; \propto \; \sigma_R(R)$. Van der Kruit (1988) studied the
vertical velocity
dispersions in external spirals and found the additional relation
${\sigma_z}(z) =
\sigma_{z}(0)\exp\left[ - {z}/{h_z} \right]$, where $\sigma_z$ increases by
a factor of 1.3 at 500 pc above the midplane. We adopt this relation, together
with
slopes of $d({\mbox{log}}\sigma_{\phi})/d{R} \; = \; 0.064$ km s$^{-1}$
kpc$^{-1}$
and $d({\mbox{log}}\sigma_{R})/d{R} \; = \; 0.051$ km s$^{-1}$ kpc$^{-1}$ from
Lewis \& Freeman (1989). In this case, $\sigma_{{\phi},z}$ are both given
by
\begin{equation}
	{\mbox {log}}\sigma_{{\phi},z}(x) = {\mbox{log}}\sigma_{{\phi},z}(0) + 0.065x
\label{sigmaphi}
\end{equation}
where $x$ is distance measured inward from the solar circle in kpc,
$\sigma_{\phi}(0) \:=\: 20$ km s$^{-1}$, and $\sigma_z(0) \:=\: 17$ km
s$^{-1}$.
Azimuthal drift is typically of order $1/5$ the dispersion along the line
of sight (Merrifield 1992) and is neglected.

The main peak of the observed microlensing timescales can be accounted for by
bulge self-lensing (ZRS), so
unlike the case for the bulge, the data do not place strong constraints on the
disk MF. If the assumed
bulge MF is also adopted for the disk, the resulting duration distribution is
consistent
with MACHO and OGLE observations in the sense of fitting the main peak (see ZRS
Figure 2). However,
the present-day MF of the disk is believed
to follow a power-law behavior which flattens out around sub-solar masses
(Scalo 1986, Larson 1986). Scalo
(1986) finds a flattening in the MF below $1.0 \Msolar$ with a maximum near
$0.3 \Msolar$, while
HST observations by Gould, Bahcall, \& Flynn (1995) find a maximum near $0.5
\Msolar$.
Furthermore, the line of sight to Baade's Window reaches a height of
350 pc above the plane at 5 kpc, so the most massive stars with typical
$\sigma_z = 10$ km s$^{-1}$
and lifetimes $<$ 10${^8}$ yr will generally not be seen along the line of
sight (B3 stars, with a lifetime of $2 \times 10^7$ yr can only be expected to
reach a height of
200 pc). However, B5 stars, with a lifetime of $2 \times 10^7$ yr, may
typically reach 600 pc (Mihalas \& Binney 1982), and
A stars are observed by Rodgers \& Harding (1989) at latitudes of
$b=25^{\circ}$. Thus, we will also consider the HST MF,
$\log{\phi}=1.35-1.34\log(\mass/{\Msolar})-1.85\left[\log(\mass/{\Msolar})\right]^{2}$ in the range from $0.1 \Msolar \leq {\mass_l} \leq 5.0 \Msolar$, as well as a composite power-law
MF with $dN/d\mass = 0$ for $0.1 {\Msolar}
\leq {\mass_l} \leq 0.4 \Msolar$, $dN/d\mass \; {\propto} \; \mass^{-1.25}$ for
$0.4 {\Msolar} \leq {\mass_l} \leq 1.0 \Msolar$, $dN/d\mass \; {\propto} \;
\mass^{-2.3}$ for
$1.0 {\Msolar} \leq {\mass_l} \leq 3.0 \Msolar$ and $dN/d\mass \; {\propto} \;
\mass^{-3.2}$ for
$3.0 {\Msolar} \leq {\mass_l} \leq 5.0 \Msolar$ (Mihalas \& Binney 1982).
Presumably, any nearby disk lenses near
or above our upper mass cutoff would be easily identifiable.

\subsection{Mass-Luminosity and Color-Color Relations}

Absolute $V$ and $K$ magnitudes for main-sequence stars in the range $0.1
{\Msolar} \leq {\mass_l} \leq
1.2 {\Msolar}$ (whether sources or lenses) were calculated using the
mass-luminosity
relations found by Henry \& McCarthy (1993) for the lower main sequence. The
CMD of stars in Baade's Window indicates a main-sequence turnoff point near 1
$\Msolar$
with a systematic brightening, at nearly constant color, of stars near the
turnoff that
have undergone some post-main-sequence
evolution (Vandenberg \& Laskarides 1987). Thus, bulge stars more
massive than 0.9 $\Msolar$ were assumed to be semi-evolved
and given an increase in magnitude between 0 (for 0.9 $\Msolar$) and 1 (for 1.2
$\Msolar$).
$B$ and $V$ magnitudes
for disk stars more massive than 1.2 $\Msolar$ were obtained using a $10th$
order fit to
mass-luminosity data from Allen (1973). Sources corresponding to clump giants
were assumed
to have absolute magnitudes of 0.335 and 1.43 in the $I$ and $V$ bands,
respectively.
All stellar magnitudes in other bands were obtained using polynomial
color-color
relations derived from accurate photometric data (Caldwell et al.~1993). The
distribution
of source magnitudes and colors obtained with this model resembles the OGLE
data (Udalski et al.~1994a).

A simple model
for reddening was adopted wherein the dust is confined to a narrow layer
centered on the Galactic midplane. The extinction in each waveband,
$a_\lambda$,
is thus assumed to increase linearly with distance out to 2 kpc, beyond which
the line of
sight to Baade's Window breaks free of the dust layer (Arp 1965), and all
sources and lenses are
assumed to suffer maximal extinction, according to the standard extinction
curve given by Savage \& Mathis (1979).

It remains to relate $r_\lambda$ to the apparent luminosities and
apparent magnitudes. Since
\begin{equation}
     m_\lambda = -2.5 \log {\ell_{\lambda}} + \mbox{constant},
\label{appmag}
\end{equation}
where $m_\lambda$ is the apparent magnitude in waveband $\lambda$, we can write
$r_\lambda$ as
\begin{equation}
     r_\lambda= {\ell_{s\lambda} \over \ell_{s\lambda} + \ell_{l\lambda}}
             = \left( 1+10^{[(m_{s\lambda} + a_{s\lambda}) - (m_{l\lambda} +
a_{l\lambda})]/2.5} \right)^{-1}.
\label{rlambda}
\end{equation}

\section{TIMESCALE DISTRIBUTION}

The optical depth is defined to be the probability that any
given star falls within the Einstein ring (of cross section $\pi
R_e^2$)  of a lens.  For a source at a distance $D_{os}$, the
optical depth is
\begin{equation}
     \tau(D_{os})=\int_0^{D_{os}} \, dD_{ol}\, n_{l}(D_{ol})
     \pi R_e^2,
\label{firsttau}
\end{equation}
where $n_{l}(D_{ol})$ is the number of lenses per unit
volume at a distance $D_{ol}$ from the observer.
Using $R_e^2=4G\mass_{d}D/c^2$, where
$D\equiv(D_{os}-D_{ol})D_{ol}/D_{os}$, we find
\begin{equation}
     \tau(D_{os})={4 \pi G \over c^2} \int\, dD_{ol} \rho_{
     l}(D_{ol}) D,
\label{secondtau}
\end{equation}
where $\rho_{l}$ is the mass of lenses per unit volume
along the line of sight.  For a fixed lens mass density,
the optical depth is independent of the mass.

The number of source stars undergoing microlensing (i.e., which
fall behind the Einstein ring of a lens) at any given time is
\begin{equation}
     N={4\pi G \over c^2} \int_0^\infty\, \tau(D_{os}) n_{s}(D_{os}) dD_{os},
\label{totalnumber}
\end{equation}
where $n_{s}(D_{os})$ is the number of sources per unit
length (in contrast to $n_{l}$, which is a number per
unit volume) along the line of sight.  So we may write
\begin{equation}
     N={4\pi G \over c^2} \int_0^\infty\, dD_{os}\, n_{s}
     (D_{os})\, \int_0^{D_{os}}\, dD_{ol}\, \int\, d{\mass_l}
     {d\rho_{l} \over d{\mass_l}}\, \int\, dv\,
     f(v,D_{ol},D_{os}).
\label{almostfinalnumber}
\end{equation}
Here, we have included the expression for $\tau(D_{os})$.
Furthermore, we have included an integral over the mass
distribution $d \rho_{l} / d{\mass_l}$ (where ${\mass_l}$ is the lens
mass) of the lens mass density,
and an integral over the
velocity distribution function $f(v,D_{ol},D_{os})$.  The
quantity $v$ is the transverse speed of the lens through the
microlensing tube, so the velocity distribution function will in
general depend on the lens and source distances.  If the mass
function of lenses is independent of distance, then the
distribution may be written
\begin{equation}
     {d\rho_{l} \over d{\mass_l}} = \rho_{l}(D_{ol})\, {1
     \over \VEV{{\mass_l}}}\, {d  {\cal N} \over d{\mass_l}}
\label{drhoequation}
\end{equation}
where $d{\cal N}/d{\mass_l}$ is the lens lens mass function and
$\VEV{{\mass_l}}$ is the mean lens mass.  The velocity and mass
distribution functions are normalized such that
\begin{equation}
     \int\,{\mass_l}\,{d{\cal N} \over d{\mass_l}}
d{\mass_l}=\VEV{{\mass_l}}\qquad {\rm and}
     \qquad \int\,f(v,D_{os},D_{ol}) dv=1.
\label{distributionsnormalization}
\end{equation}
Therefore, Eq.~(\ref{almostfinalnumber}) can be written,
\begin{equation}
     N={4\pi G \over c^2 \VEV{{\mass_l}}} \int_0^\infty\, dD_{os}\, n_{s}
     (D_{os})\, \int_0^{D_{os}}\, dD_{ol}\,
     \rho_{l}(D_{ol})\, \int \, dm\,m {d{\cal N} \over
     dm} \int\, dv\, f(v,D_{ol},D_{os}).
\label{finalnumber}
\end{equation}
The total number of sources undergoing microlensing at any given
time is independent of the lens-mass and velocity distributions,
so the integrals in Eq.~(\ref{finalnumber}) evaluate trivially
to unity.  They are included for use in the next step.

The distribution of timescales $t_0=R_e/v$ for all events in
progress at any given time is
\begin{eqnarray}
     {dN \over dt_0} & = & {4\pi G \over c^2 \VEV{{\mass_l}}}
     \Bigg[\int_0^\infty\, dD_{os}\, n_{
     s}(D_{os})\, \int_0^{D_{os}}\, dD_{ol}\, \rho_{d
     }(D_{ol}) D \nonumber \\
     		     &   & \int\, d{\mass_l}\, {\mass_l}{d {\cal N} \over
d{\mass_l}} \, \int\, dv\,
     f(v,D_{ol},D_{os})\, \delta(t_0-\sqrt{4G{\mass_l}D}/(cv))\Bigg].
\label{numberdistribution}
\end{eqnarray}
The frequency of events with time scales in the range $t_0$ to
$t_0+dt_0$ is
\begin{equation}
     {d\Gamma \over dt_0} dt_0 = {2\over \pi t_0} {dN \over
     dt_0} dt_0,
\label{Gammadefinition}
\end{equation}
so\footnote{We thank A. Gould for a discussion of the timescale
distribution.}
\begin{eqnarray}
     {d\Gamma \over dt_0} & = & {4 G^{1/2} \over c \VEV{{\mass_l}}}
     \Bigg[\int_0^\infty\, dD_{os}\, n_{
     s}(D_{os})\, \int_0^{D_{os}}\, dD_{ol}\, \rho_{d
     }(D_{ol}) D^{1/2} \nonumber \\
                          &   & \int\, d{\mass_l}\, {\mass_l}^{1/2} {d
{\cal N} \over
     d{\mass_l}} \, \int\, dv\,
     v f(v,D_{ol},D_{os})\, \delta(t_0-\sqrt{4G{\mass_l}D}/(cv))\Bigg].
\label{Gammaequation}
\end{eqnarray}
Due to the presence of the $\delta$ function in
Eq.~(\ref{numberdistribution}), the prefactor $t_0^{-1}$ can be
taken inside the integral and evaluated to
$t_0=\sqrt{4G{\mass_l}D}/(cv)$.  The expression for the total frequency
$\Gamma$ is obtained by removing the $\delta$ function in this
expression.

The total number of events which are observed to peak within any
given observing period of duration $\Delta t$ (assumed to be
long compared with $t_0$) is $N_{\rm peak}=\Gamma \Delta t$, so
the observed timescale distribution is
\begin{equation}
     f(t_0) \propto {d\Gamma \over dt_0} {\epsilon}(t_0).
\label{durationdistribution}
\end{equation}
where we have explicitly included the observation detection efficiency
${\epsilon}(t_0)$.  Following ZSR, we adopt the function
${\epsilon}(t_0)=0.3\exp[-(t_0/11 \mbox{ days})^{-0.7}]$ as a good
interpolation of the values given in Udalski et al.~(1994) for the OGLE data.
The MACHO efficiencies
are typically higher for a given event duration (Alcock et al.~1995b), but do
not significantly alter our
results for color shifting.

The MACHO and OGLE surveys, which sample events roughly daily
with typical errors of roughly 0.05 mag, do not constrain the microlensing
light curves
sufficiently to pick out color shifting in most cases. Thus, the condition that
events be achromatic should have little effect on the observed timescale
distribution
for events of short duration (i.e., $t_0 \leq 20$ days). However,
longer-duration events,
with better sampled light curves, are expected to yield a greater fraction of
CSEs,
particularly for the case of disk lensing (see Sections 7 and 8, and Figure
10).
If the achromaticity condition is strong, it will then have the effect of
systematically eliminating larger fractions of the longer events, so that the
contribution of these events to the overall duration distribution
would be suppressed relative to the ``actual'' distribution. This effect
becomes
more dramatic as the sampling interval and errors are reduced. If most of the
lenses
are in the form of main-sequence stars, this leads to the danger that the
resulting
color shifts prevent the detection of those events for which the degeneracy in
$t_0$ can
be broken entirely. However, the current MACHO collaboration cut on
achromaticity
is sufficiently weak that this is unlikely to be a problem (Griest 1995). In
particular, we
estimate that this effect is negligible for the case of bulge self-lensing,
which accounts for most of the
observed events, and is only expected to shift the disk lensing peaks by about
10\% towards
the shorter end.

The duration distributions for the disk and bulge models used are
shown in Figure 3. The relative peak heights have not
been normalized. ZRS fits the Zhao bar model
to 55 MACHO and OGLE events and finds that the best-fit model,
with a median lens mass of
$0.18 \Msolar$, predicts 40 bar events with a mean duration of 16 days, and 14
disk
events with a mean duration of 30 days. Our bar model, with a median lens mass
of $0.185\Msolar$
is nearly identical to that of ZRS and fits
the main peak of the observed MACHO and OGLE distributions (see ZRS, Figure 2).
\begin{figure}
\plotone{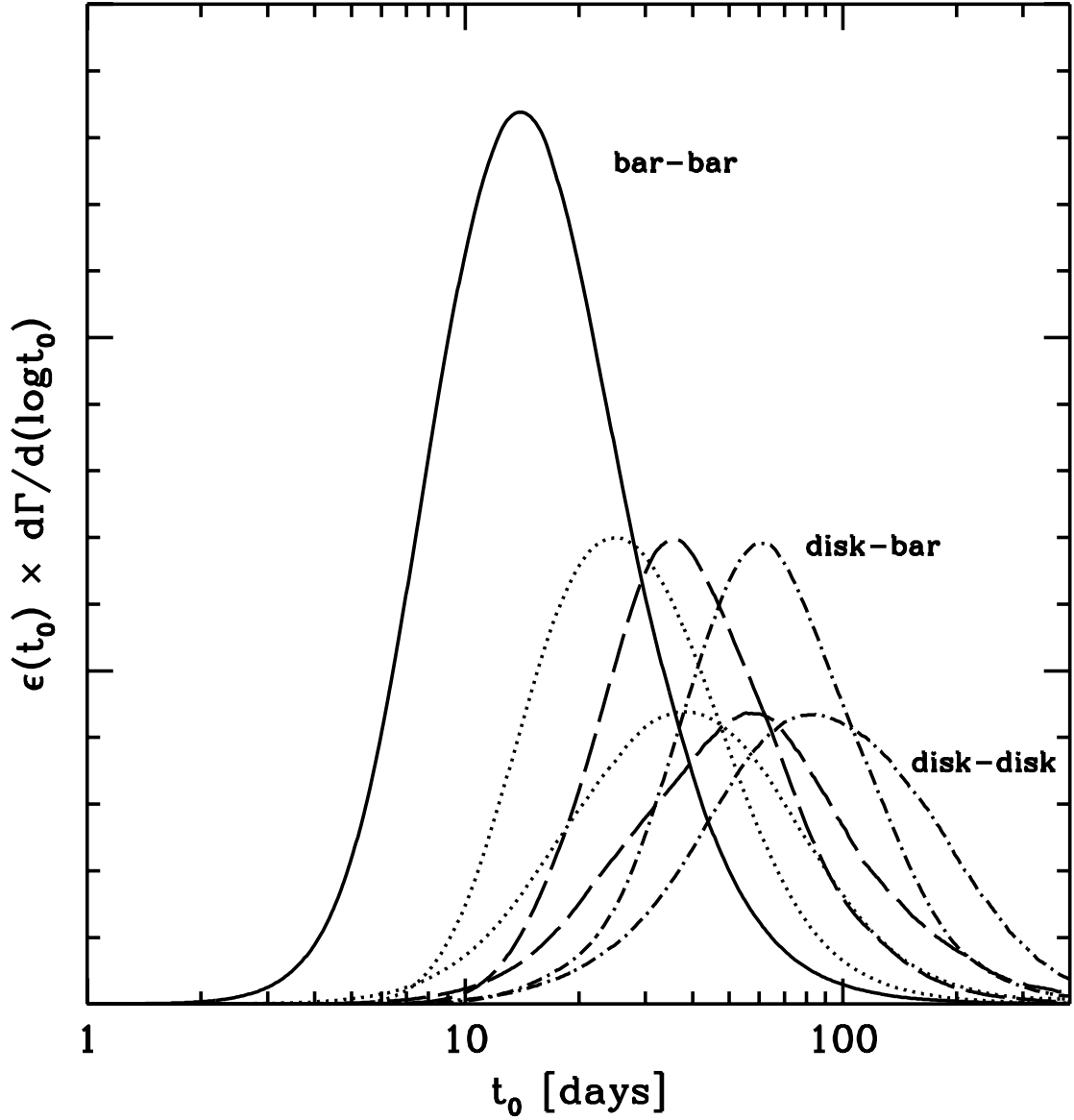}
\caption{Predicted microlensing event duration distributions (non-normalized)
for the bulge and disk models used. The main peak
due to bulge self-lensing matches the peak of the observed MACHO and OGLE
distributions. The dotted,
dashed and dot-dashed curves represent the logarithmic, HST, and composite
power-law disk MFs, respectively. We speculate
that the disk lensing events may account for the few long-duration events
observed.}
\label{timescale1}
\end{figure}

The distributions for the various MF models differ considerably. ZRS finds that
the
logarithmic disk MF is consistent with current microlensing data, but this
MF differs considerably from that of the solar neighborhood. The HST and
composite power-law MFs, derived from observations of the disk, produce peak
durations near 40 and 63 days respectively,
due to events where disk stars lens bulge stars. These results are less
consistent
with the main peak of the MACHO and OGLE distribution. However, both Han and
Gould (1995b, see their Figure 3) and ZRS
find significant fractions of observed events with $t_0 \geq 60$ days,
including an apparent
bump in the duration distribution near $t_0=100 \mbox{ days}$, which are not
accounted for
by their models. The longer events in particular seem to be systematically
located closer to the Galactic plane
than the rest of the events (Zhao, private communication; see Figure 1 of
Bennett et al.~1995), which suggests
a disk origin. Our disk models using the HST and power-law MFs predict
significant
contributions from disk-lensing-bulge events at long durations, particularly by
the latter.
Furthermore, both models predict broad peaks near 60 and 90 days, respectively,
due to disk self-lensing.
Mollerach \& Roulet (1995) also predict a substantial number of long-term
events arising from disk lensing.
Therefore, we speculate that the excess of long-duration events arise from
disk-bulge and disk-disk lensing, with the lower amplitudes simply
corresponding to the smaller optical depth of the disk.
In this scenario, the larger timescales would probably arise from massive
lenses
with low velocity dispersions, i.e., nearby bright stars, but no conspicuously
bright lenses have been identified.
It is likely, however, that color-shift analysis should help to constrain the
nature of these long-term events.

Another explanation for the long-term events would be a distribution of sources
{\em and} lenses
associated with the so-called ``young'' disk, having mean
velocity dispersions of 10 km s$^{-1}$ characteristic of the gas layer. A
low-mass population (e.g.,
following the logarithmic disk MF) with these characteristics would also give
rise to a broad disk self-lensing peak near
$t_0=100$ days (although this would require more mass in the young disk than
usually assumed), while the higher-dispersion population would remain
consistent with the main peak.\footnote{We thank
R. Olling for this suggestion.}
It is currently believed that the young disk accounts for 20\% of the
disk light (van der Kruit 1990). Han and Gould (1995b) point out that a similar
result is
obtained assuming the existence of a disk population of black holes, white
dwarfs, and
neutron stars which have low dispersion. Improved microlensing statistics will
help to constrain better the
MF of the lens population. In calculating the expected rates for CSEs, we
conservatively adopt the logarithmic
MF for the disk, noting that the low median lens mass in this model implies a
lower limit on the color-shifting rate for
disk events.

\section{SENSITIVITY TO SIGNAL}

Given the bar and disk models above, we wish to investigate the
fraction of events that give rise to events which have
observable shape distortions and/or color shifts.

As explained above, the amplification observed during a
microlensing event in waveband $\lambda$ is given by
Eq. (\ref{observedamplification}) where $A(t)$ is the
microlensing amplification in Eq. (\ref{einsteinradius}).
If $A_T$ is set, there are three parameters, $t_{\rm max}$, $A_{\rm max}$, and
$t_0$,
which are needed to specify $A(t)$.  There is then an additional
parameter, $r_\lambda$, the luminosity offset ratio, needed to
describe the light curve, $\Ampl_\lambda(t)$.  Therefore,
color-shifted light curves in a number $n$ of different
wavebands are fit by $3+n$ parameters.  (The baseline
brightnesses in the $n$ wavebands must also be fit; however,
here we assume that the long-baseline observations of the source
stars before and after the event will assure that the baselines
brightnesses will be measured to great precision.)

We will consider frequently sampled followup observations
of microlensing events and investigate the sensitivity of such a
program to shape distortions in the microlensing light curves.
We parameterize the various possible observing strategies first
by the frequency of sampling, so measurements of the
brightnesses in each waveband are obtained at times $t_i$
separated by a time interval $\Delta t$.  We also consider different
photometric accuracies.  Since the change in magnitude and the
amplification are related by ${\Delta}m = 2.5{\mbox{ log}A}$, a constant
error in $m$ translates to a constant percent error in $A$.
Thus we will assume that any given
measurement of the amplification has a gaussian distribution
centered at the true value with a variance $\sigma_l\simeq
f\Ampl(t_l)$ which is some fraction $f$ of the amplification
corresponding to the magnitude error.

To determine whether the shape distortion in a given
event---specified by some source-star and lens masses and
distances---will be observable, we first calculate the expected
observed amplification $\Ampl(t)$, and then ask whether
light-curve measurements with a given sampling frequency and
fractional error will produce a value of $r_\lambda$ that is
different from unity at the 95\% confidence limit.  If one of
the unlensed-light parameters $r_\lambda$ differs from unity
with this statistical significance, then a shape distortion from
unlensed light has been observed.

Suppose that the light curves are sampled $N$ times and the error
on each measurement $i$ is gaussian distributed about the mean
with a variance $\sigma_i$, as discussed above.  Any given
light curve $\Ampl(t,{\bf s})$ will depend on the set of
parameters ${\bf s} =
\{A_{\rm max}, t_{\rm max}, {t_0}, r_\lambda\}$ (for
$\lambda=1,..,n$).  If the predicted parameters for a given
event are ${\bf s_0}$, then the probability
distribution for observing a light curve which is best fit by
the parameters ${\bf s}$ is (Gould 1995b; Jungman et al.~1995)
\begin{equation}
P({\bf s}) \propto \exp\left[ -{1\over 2}({\bf s}-{\bf s}_0)
                     \cdot [\alpha] \cdot({\bf s}-{\bf s}_0)\right],
\label{likelihood}
\end{equation}
where the curvature matrix $[\alpha]$ is given approximately by
\begin{equation}
\alpha_{ij} = \sum_{l=1}^{nN} {1\over\sigma_l^2}
              \left[{\partial \Ampl(t_l,{\bf s}_0)\over\partial s_i}
                    {\partial \Ampl(t_l,{\bf s}_0)\over\partial
		    s_j}\right],
\label{curvature}
\end{equation}
where the sum is over each measurement of the amplification in
every waveband observed.  The covariance matrix is given by
$[{\cal C}] = [\alpha]^{-1}$.  It is an estimate of the
matrix of standard errors that could be obtained by a fit to
data and is used to define the elliptical boundary of a desired
confidence region. The projections of the ellipse define the corresponding
confidence interval $\delta{\bf s}$ on the parameters of interest.
For our purposes, if the ``true''
value of the unlensed-light parameter is $r_\lambda$, the
probability of observing a light curve which is best fit by a
value $r_\lambda^{\rm obs}$ is a gaussian centered on
$r_\lambda$ with a variance ${\delta{r_{\lambda}}(68.3\%)}$, which
corresponds to the $1{\sigma}$ confidence level.
As we cycle through each event, if $\delta{r_{\lambda}}(95\%) < (1-r)$
for any $\lambda$, then the given event will
have a shape distortion that can be distinguished with 95\%
confidence, and we accept this as a shape-distorted event.
Otherwise, the event cannot be distinguished from a standard
achromatic microlensing light curve. Our simulation included only those events
in which
the source star had an apparent magnitude of $V < 21.0$, and the event duration
$t_e$ satisfied
2 days $\leq t_e \leq$ 1 year. All included events were weighted
by their frequency as given by Eq.~(\ref{Gammaequation}) and summed to give
an overall normalization. The fraction of color-shifted events, $F_{CSE}$, is
then defined
as the ratio of the weighted sum of all color-shifted events to the
overall normalization.

\section{RESULTS}
\subsection{Accessible Masses}

Before turning to our calculation of expected rates for CSEs, we
determine, for a typical event, the values of $\mass_l$ (as a function
of $D_{ol}$) which will produce an observable color shift, considering
separately
the cases of main-sequence and clump-giant sources in the bulge.
Given identical achromatic light curves in two wavebands $\lambda_1$ and
$\lambda_2$, each consisting
of $N$ points with normally distributed errors $\sigma_{1}^{(i)}$ and
$\sigma_{2}^{(i)}$,
we can find the minimum values of $r_{{\lambda}_1}$ and $r_{{\lambda}_2}$
consistent with
unity (i.e., achromaticity) at the 95\% confidence level. In other words we
look for the
maximum possible $\ell_{d{\lambda}}$ consistent with the assumption of
achromaticity ($\ell_{s{\lambda}}$ is assumed
to be known). Together with an assumed mass-luminosity relationship, this
places an upper
limit on the lens mass as a function of distance, at the desired confidence
level. This upper
limit can be established from observations in a single waveband. However,
multiple wavebands provide
additional independent limits and thus tighter constraints.
Figures 4 and 5 display upper limits on $\mass_l$ as function of $D_{ol}$, from
observations in different wavebands,
for typical events with main-sequence and giant sources at 8 kpc, respectively.
We assume an event
timescale of $t_0 = 15$ days, near
the peak of the bulge self-lensing duration distribution, and a minimum impact
parameter of $u_{\rm min} = 0.2$.
Any lens mass above the curves will give rise to a detectable
color shift at the 95\% confidence level, for the observational parameters
listed in the Figure caption.
\begin{figure}
\plotone{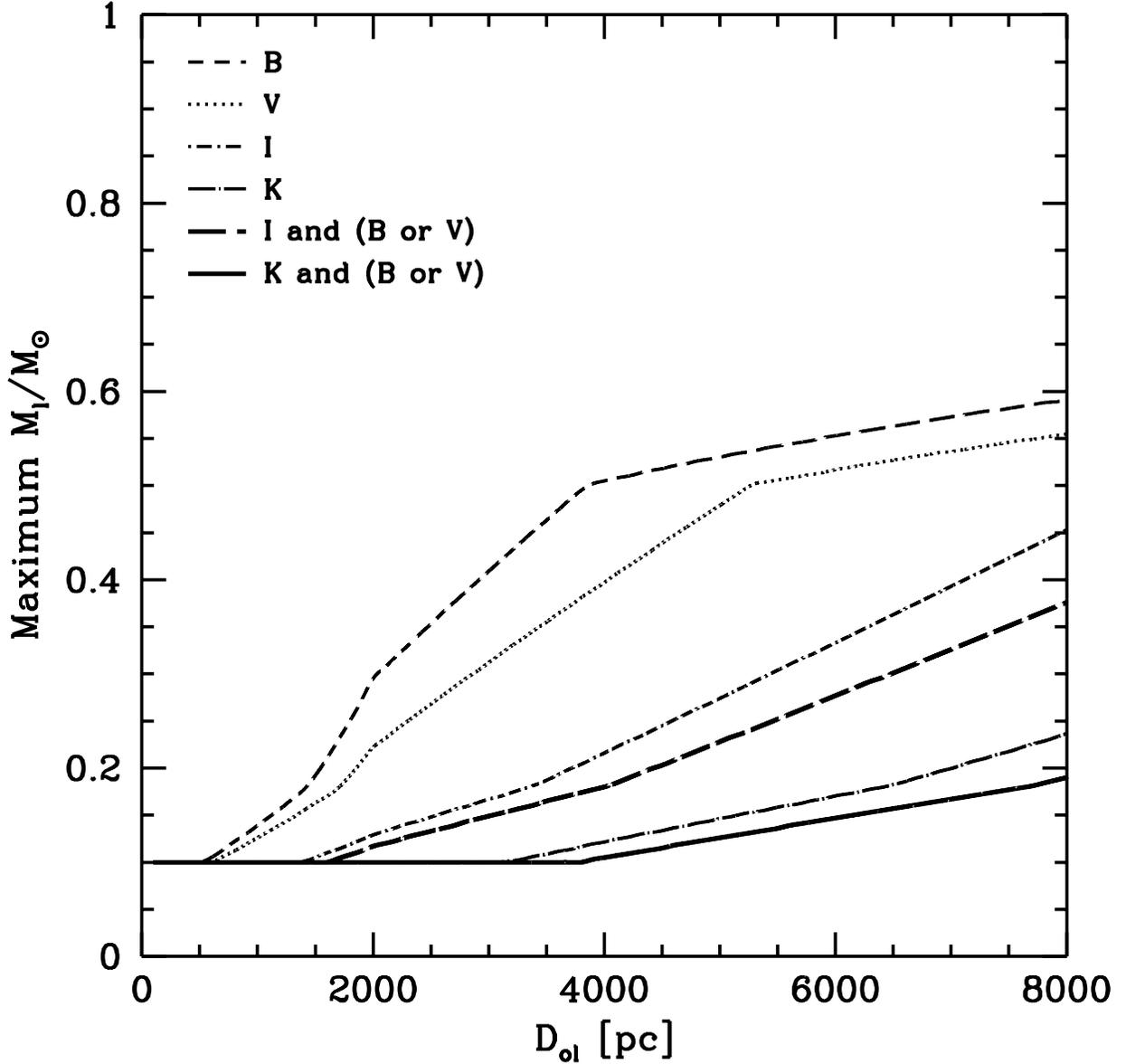}
\caption{Upper limits on the lens mass, $\mass_l$, as a function of
observer-lens distance, $D_{ol}$,
at the 95\% confidence level for
a model event with $D_{os} = 8$ kpc, $t_0 = 15$ days, $u_{\rm min} = 0.2$ and
simulated
data with 0.01-mag normally distributed errors, taken hourly while $A >
A_T=1.34$. The source star
is a $1.0 \Msolar$ main-sequence star. Note that the strongest constraints come
from $I$ and $K$,
and in particular that for events observed in $K$ and either $B$ or $V$, all
lenses toward the bulge
with $\mass_l \geq 1.8 \Msolar$ will give rise to a color shift.
All results were obtained using the mass-luminosity and color-color relations
stated above.
The kinks are due to the tri-modal nature of the Henry \& McCarthy (1993)
mass-luminosity relationships.}
\label{masslimms}
\end{figure}
\begin{figure}
\plotone{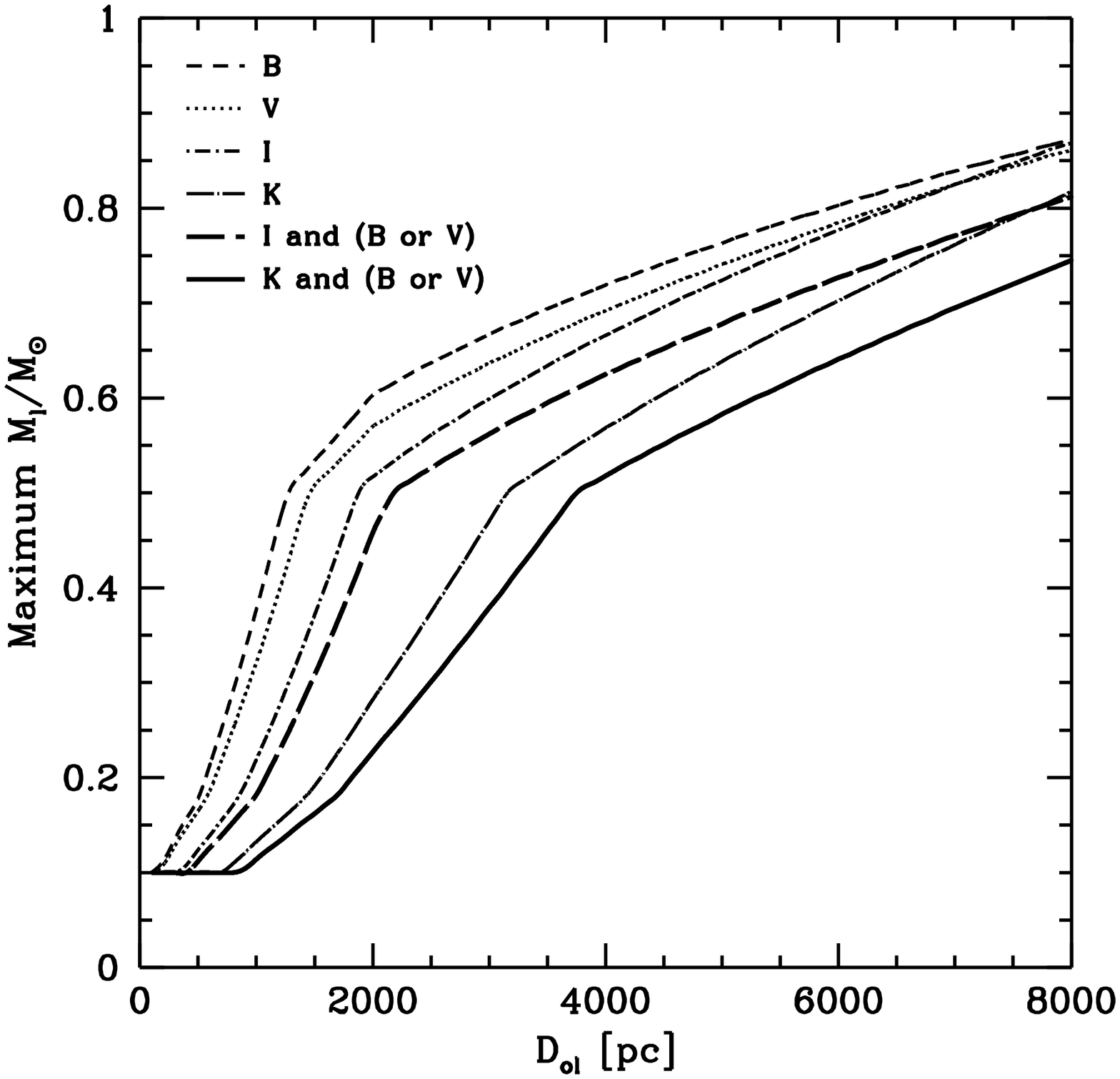}
\caption{Same as Figure 4, with a clump-giant source at 8 kpc. The constraints
are weaker due to
the smaller fractional contribution of the faint lens to the amplification of
the brighter source.}
\label{masslimgiant}
\end{figure}

The results show that redder wavebands, where the lenses are
assumed to emit most of their light, yield the strongest constraints on the
lens mass all
the way down to our lower cutoff of $0.1 \Msolar$. In particular, we find that
for events with main-sequence sources, observed
in $K$ and either $B$ or $V$, any lenses with $\mass_l > 0.18 \Msolar$ and all
lenses in our
model closer
than 4 kpc in general, will give rise to a measurable color shift
at the 95\% confidence level. The results in the bluer wavebands, and for giant
sources in general,
are less dramatic, since the fractional contribution of the lens light will be
smaller.
The typical maximum achromatic lens masses in these cases are 0.3 to 0.6
$\Msolar$ and 0.5 to 0.8 $\Msolar$, respectively.
Of course, if the photometric errors are reduced, and/or the sampling frequency
is increased, the constraints on
the lens mass become stronger. The upper limits also decrease for larger event
timescales.

We further applied this analysis to the MACHO-collaboration parallax event
(Kamionkowski \& Buchalter 1995;
Alcock et al.~1995b), since this
event represents the most well-sampled existing microlensing light curve. To do
so, we simulated
the light curve, sampling frequency, and photometric errors of the observed
event.
Our results are shown in
Figure 6, and agree with the results of Alcock et al.~(1995b).
The weaker constraints on the lens mass, as compared to Figure 5,
suggest that surveys conducted with higher sampling frequency and greater
photometric precision will
significantly improve our knowledge of the lens MF.
\begin{figure}
\plotone{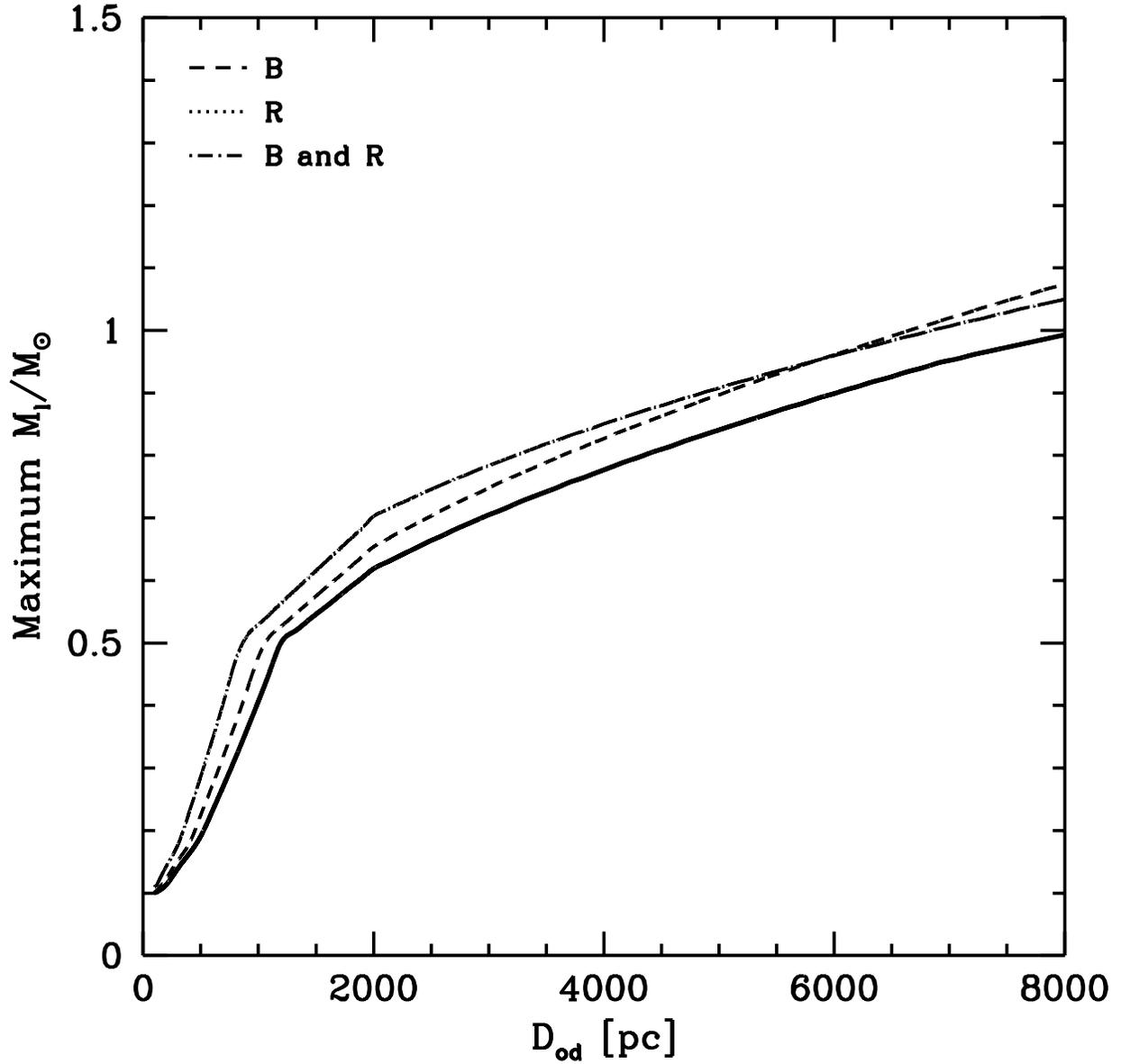}
\caption{Upper limits on $\mass_l$ as a function of $D_{ol}$ at the 95\%
confidence level for the
MACHO-collaboration parallax event, with simulated parameters $D_{os} = 8$ kpc,
$t_0 = 100$ days, $u_{\rm min} = 0.1$ and
``data'' with 0.03-mag errors, taken every 24 hours while $A > A_T=1.34$.}
\label{masslimparallax}
\end{figure}

\subsection{Color-Shifted Fractions}

Having illustrated what masses and distances will give observable distortions
of
the light curves, we now
employ our Galactic model to calculate the overall expected fractions of
color-shifted
events (i.e., for all timescales), as functions of sampling frequency and
photometric error.
We distinguish
between surveys focussed on source stars which are giants (denoted G) and those
focussed on
main-sequence sources (denoted MS). Despite smaller photometric errors,
giants are expected to give rise to fewer
CSEs, since the relative contribution of the faint
main-sequence lens to the overall light profile will be small compared
to the amplification of the bright giant star. We also consider
separately the cases of bulge self-lensing (BB), disk stars lensing bulge
sources (DB),
and disk self-lensing (DD), giving rise to a total of six possible scenarios.
Although
the disk lensing events contribute less to the overall optical depth
(particularly in
the case of self-lensing of the disk), the transverse speed of
disk lenses through the microlensing tube
will be characteristically lower than that of bulge lenses, giving
rise to longer-duration events.  Such events will be more heavily
sampled and thus more likely to yield a color shift. In all cases,
the full geometrical depth of the bulge and/or disk
is taken into account. We first consider observations made in $B$ and $K$,
denoted by ($B$,$K$), since these
bands are well-separated in wavelength and thus provide a large baseline for
detecting
color shifts. We also investigate the expected fractions for ($V$,$I$) to
determine
whether CCD photometry and infrared imaging are necessary for observing this
effect.
The logarithmic disk MF, which is consistent with the main timescale peak of
the observed MACHO and OGLE events, is expected
to give the fewest detectable color shifts for disk events. We examine this
case
and later compare with other MFs.

Figures 7 and 8 display the results of our Monte Carlo
simulations, showing $F_{CSE}$, the fraction of events which are
observably distorted, as a function of sampling interval,
for events observed in ($B$,$K$) and ($V$,$I$), respectively, for the
various configurations. The solid, short-dashed, and long-dashed curves
represent photometric errors of 0.05, 0.01 and 0.001 mag, respectively.
The results verify that existing data on microlensing towards the
bulge, with typical photometric errors of up to 0.05 mag and daily sampling
rates at best, is not expected to be sensitive to these small distortions. Even
for the most favorable
lens-source configurations [namely DB(MS) and DD(MS)], current surveys, which
use ($V$,$R$) or ($V$,$I$), should only observe
a color shift in fewer than 4\% of the events; for the most likely scenario,
BB(MS), the expectation is less than 2\%.
However, the Figures indicate significant increases in $F_{CSE}$ for sampling
intervals less than 5 hours, and particularly
for intervals less than 1 hour. Furthermore, $F_{CSE}$ increases dramatically
as the errors are reduced. The
intensive microlensing monitoring programs which have been proposed incorporate
both of these improvements, and should
reveal a much higher incidence of color shifting.
\begin{figure}
\plotone{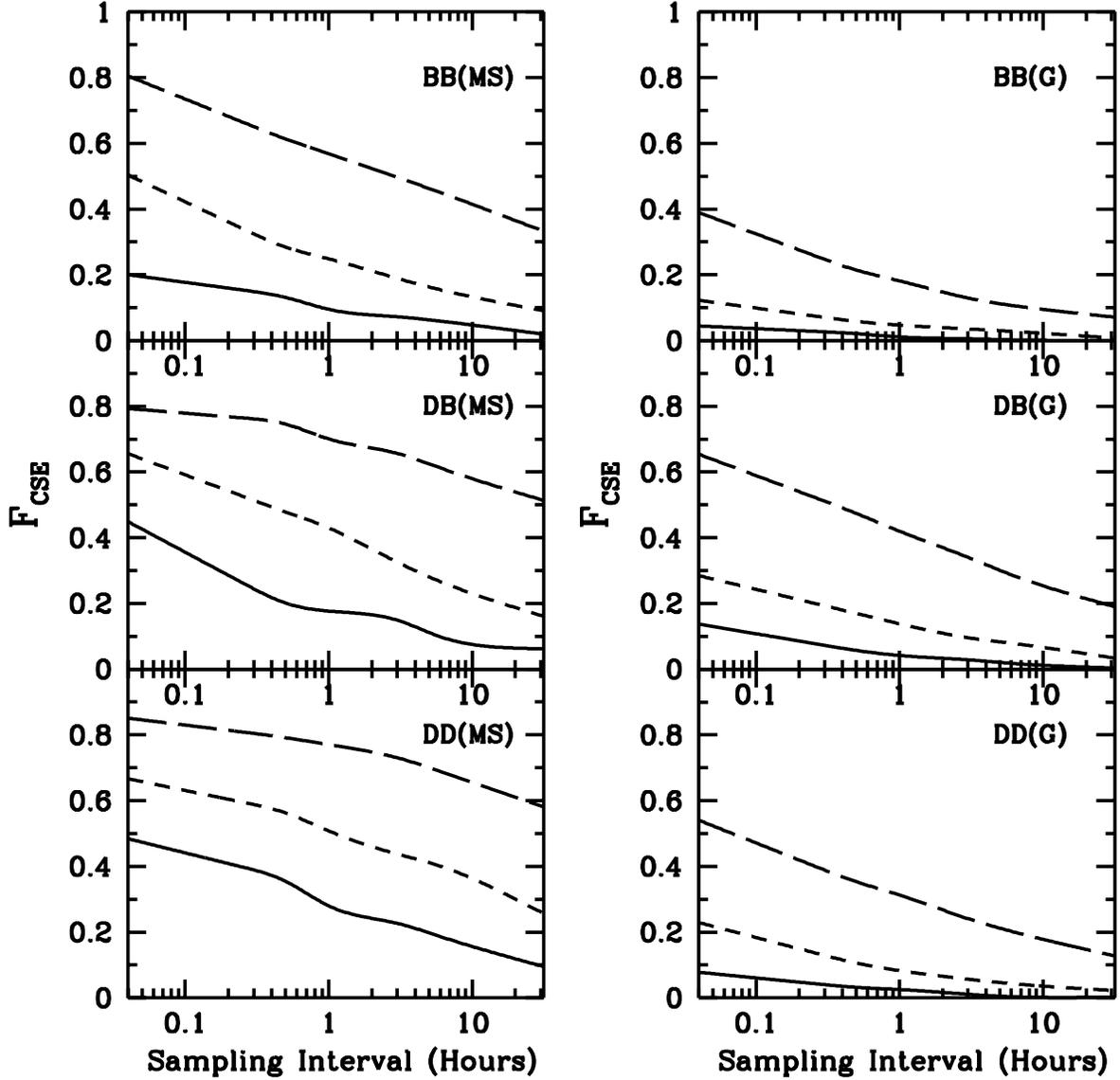}
\caption{Expected fractions of CSEs as a function of sampling interval for
events observed
in $B$ and $K$. The solid, short-dashed, and long-dashed lines correspond to
photometric
errors of 0.05, 0.01, and 0.001 mag, respectively. The various configurations
are denoted by
BB for bulge self-lensing, DB for disk-lensing-bulge events, and DD for disk
self-lensing. MS indicates
that the sources are main-sequence stars, while G indicates clump-giant
sources.}
\end{figure}

Figure 7 shows that for {\em all} cases of lens-source pairings
at least a few percent of the events will be observably distorted,
given frequent sampling and good photometry in $B$ and $K$.
For the most likely scenario of a bulge
star lensing a bulge main-sequence star [BB(MS)], we find the expected fraction
of color-shifted events is $F_{CSE} = 0.25$, given 0.01-mag photometric errors
and an hourly sampling rate. In the more favorable case of a disk star lensing
a main-sequence
bulge star [DB(MS)], we find $F_{CSE}=0.43$ given the same observational
parameters.
Since the disk optical depth to microlensing is roughly 1/5 that of the bulge
(ZSR), we estimate that 28\% of the events where a main-sequence
bulge star is lensed should yield a measurable color shift. For the case of a
bulge giant being lensed, the relative contributions of
$F_{CSE}=0.05$ from BB(G) and $F_{CSE}=0.14$ from DD(G) imply an overall
expected color shifting rate of 7\%. Moreover,
if the light curve is sampled every 15 minutes, the expected fractions jump to
37\% for bulge main-sequence sources and
10\% for giants.

At first glance, it might be expected that the rare disk-on-disk events, with
lower characteristic transverse speeds, would yield significantly
higher fractions of color shifts due to the increased event duration
$t_0={R_e}/v$.
However, since more disk stars come into view of the solid angle toward the
point where the disk is truncated ($dN/dD_{ol} \; {\propto} \; D_{ol}^2$), the
decrease in $v$ is offset
by a decrease in the typical lens-source
separation [i.e. the factor $D=(D_{os} - D_{ol})(D_{ol}/D_{os})$ in $R_e$],
unlike in
the disk-lensing-bar case
where the two model populations are spatially differentiated. Thus for hourly
sampling
and 0.01-mag errors, $F_{CSE} = 0.51$ for DD(MS), slightly greater
than the value for DB(MS), while DD(G) is expected
to yield $F_{CSE} = 0.08$. For 15-minute sampling, these become 60\% and 14\%,
respectively.

It is also of interest to examine the dependence of $F_{CSE}$ on the choice of
wavebands, to determine whether
CCD photometry and infrared imaging are necessary to observe this effect.
Figure 8 shows the results
of our calculation done using ($V$,$I$). We find that the expected rates remain
significantly high, ranging from 0.6 to 0.7 times
those for ($B$,$K$) in all cases. As mentioned above, the results for
intermediate waveband pairings
are determined primarily by the longest waveband used so that ($V$,$K$) and
($B$,$I$) yield virtually the same results as ($B$,$K$) and ($V$,$I$),
respectively. For all wavebands, $F_{CSE}$ can be increased by improved
statistics from multiple dedicated telescopes at different locations. For
imaging in ($V$,$I$), however, it is conceivable that high-precision photometry
could potentially reduce errors
to the millimagnitude regime, which is unlikely
in the $B$ and $K$ bands. Together with a reduced sampling interval of 15
minutes, this improvement could increase the overall likelihood of observing
color shifting of bulge sources to over 40\% for main-sequence stars and 13\%
for giants, for
events observed in ($V$,$I$). For disk self-lensing the
expected fractions can reach 60\% for dwarf sources and 20\% for giants.
\begin{figure}
\plotone{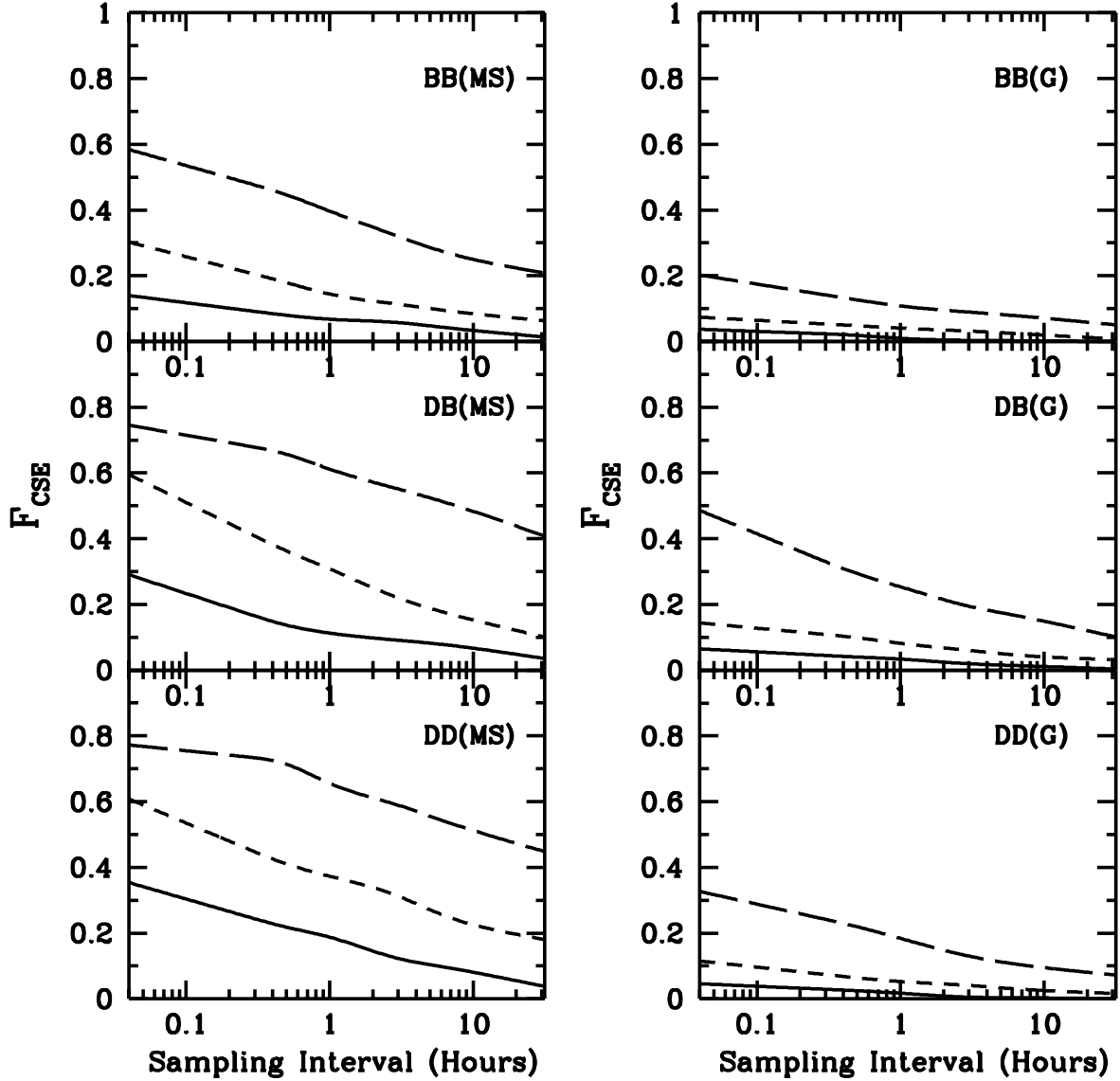}
\caption{Expected $F_{CSE}$ for events observed in $V$ and $I$. The solid,
short-dashed, and long-dashed lines correspond to photometric errors of 0.05,
0.01, and 0.001 mag, respectively. The various configurations are denoted by
BB for bulge self-lensing, DB for disk-lensing-bulge events, and DD for disk
self-lensing. MS indicates
that the sources are main-sequence stars, while G indicates clump-giant
sources. Accurate observations in $V$ and $I$ may have typical errors of
several millimagnitudes and thus yield significant increases in $F_{CSE}$.}
\end{figure}

We also note that if the stellar MF of the disk in the range between
$0.1{\Msolar} \mbox{ and }1{\Msolar}$ is in fact flatter than the assumed
logarithmic distribution, the increase in the median lens mass (and thus
luminosity)
would significantly increase the expected fraction of CSEs for cases
where the lens is a disk star. In Figure 9, we compare $F_{CSE}$ using the
logarithmic, HST, and composite power-law MFs, for
the case of disk stars lensing bulge main-sequence stars [DD(MS)], for
simulated observations in $V$ and $I$ with 0.01-mag errors.
The dramatic increase suggests that our calculations of $F_{CSE}$ for disk
events actually represent lower limits,
since it is very unlikely that the actual MF of the disk continues to rise
steeply down to $0.1 \Msolar$. Moreover, this
result indicates that color-shifted microlensing statistics for disk events can
help distinguish between the different disk models.
\begin{figure}
\plotone{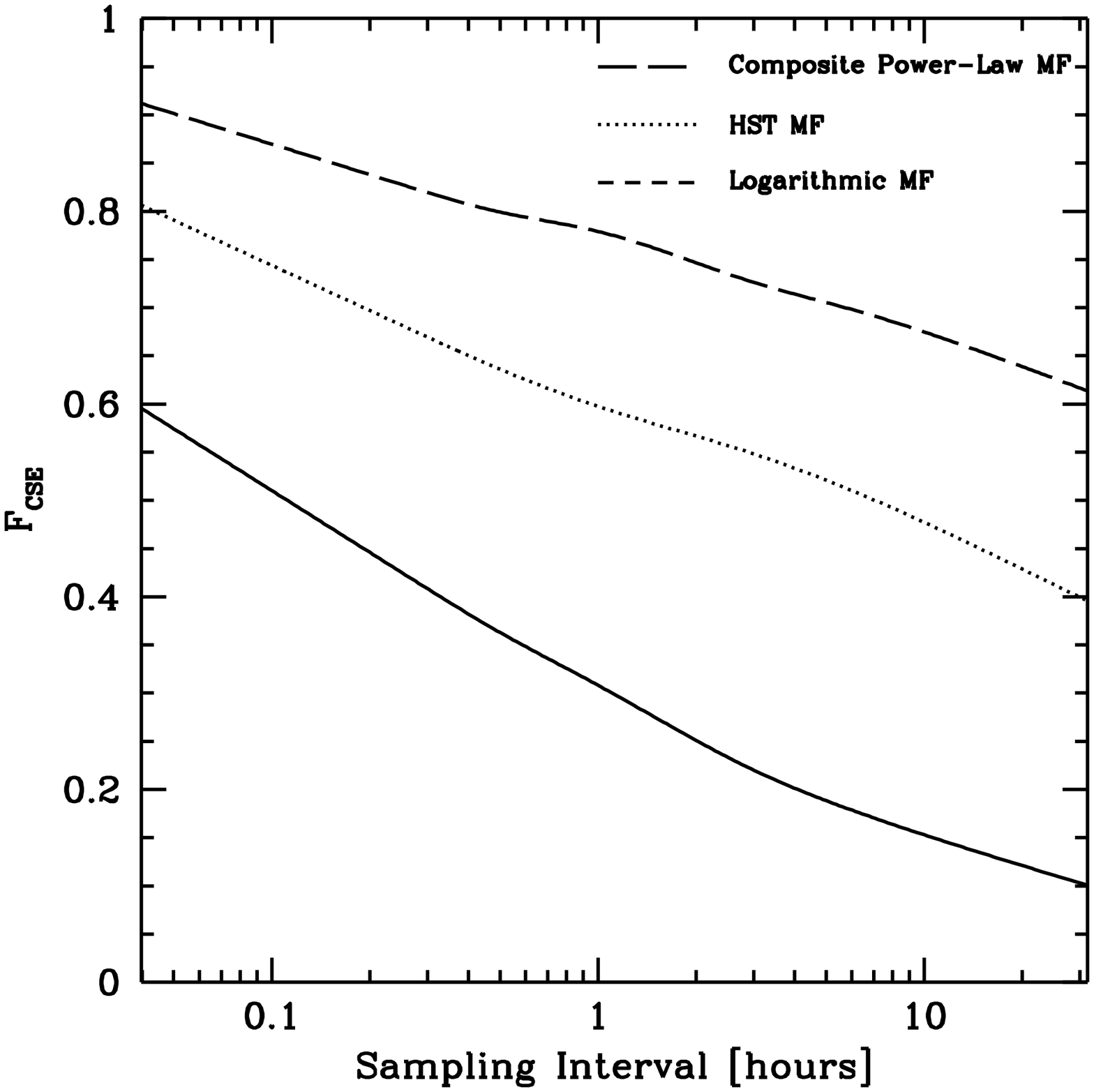}
\caption{$F_{CSE}$ vs. sampling interval for DD(MS) lensing with different disk
MFs, assuming
0.01-mag errors in ($V$,$I$). The logarithmic MF, which rises steeply to $0.1
\Msolar$, has the lowest median mass
and predicts the fewest
color-shifted events. Local observations of the disk, however, favor a MF which
flattens
below $1 \Msolar$, so the logarithmic disk MF is likely a lower limit on
$F_{CSE}$ for disk lensing.}
\end{figure}

The significant expected fractions of CSEs imply that many observed
microlensing events will
be able to provide more information than standard achromatic light curves. It
is still possible that a small
number of color-shifted events might already exist in the MACHO or OGLE data,
but would not have
been identified as such due to insufficiently precise measurement of the light
curve.
However, we restate that recent calculations of microlensing towards the bulge
(ZRS)
show that Galactic models having a mass function with 25\% or more of the mass
as
brown dwarfs are inconsistent with current data. If the lenses are low-mass
stars then color shifting will be
a real effect and the cut on achromaticity must remain weak; otherwise, it
would systematically
exclude analysis of precisely those events from which all relevant parameters
can be
disentangled. In the following Section, we outline the method for obtaining
these parameters.

\section{LIGHT-CURVE ANALYSIS}

In Section 3.1, we outlined the procedure whereby the luminosity-offset ratio
in each waveband, $r_\lambda$, can be calculated from the masses and
distances of the source and lens. Here, we follow an inverse
procedure to infer the lens properties from simulated data for a ``typical''
microlensing event, and estimate
how accurately the lens mass, distance, and transverse speed can be
reconstructed from
analysis of color-shifted light curves. In the following, we shall assume that
observations of the source can be carried out for a sufficiently long
time after the event so that the source mass and distance can be determined.
We then fit measurements in two wavebands for the two unknown
quantities $\mass_l$ and $D_{ol}$. Additional wavebands can be used, in
principle, to make the determination more precise.
For an assumed mass-luminosity relationship, Eqs.~(\ref{appmag}) and
{}~(\ref{rlambda}) can be written
\begin{equation}
r_{{\lambda}_1}=r_{{\lambda}_1}(\mass_l,D_{ol}), \qquad
r_{{\lambda}_2}=r_{{\lambda}_2}(\mass_l,D_{ol}).
\label{rs}
\end{equation}
 Equations ~(\ref{rs}) can be inverted to give
\begin{equation}
\mass_l=\mass_l(r_{{\lambda}_1},r_{{\lambda}_2}), \qquad
D_{ol}=D_{ol}(r_{{\lambda}_1},r_{{\lambda}_2}).
\label{m,d}
\end{equation}
For a given event observed in two bands, $\lambda_{1}$ and $\lambda_{2}$, the
light
curve can be fit by the five parameters ${\bf s} =
\{A_{\rm max}, t_{\rm max}, {t_0}, r_{\lambda_{1}}, r_{\lambda_{2}}\}$.
The values for $\mass_l$ and $D_{ol}$ are by obtained using Eqs.~(\ref{m,d})
with
the best-fit values for $r_{{\lambda}_1}$ and $r_{{\lambda}_2}$. We find that
the covariance matrix for a typical CSE
indicates that the errors in the $r_{\lambda}$ are typically much larger than
the errors
in $A_{\rm max}$, $t_{\rm max}$, and $t_0$. Thus, to a good approximation, the
errors on
$\mass_l$ and $D_{ol}$ are given by
\begin{equation}
({\Delta}\mass_l)^2={\left(\frac{{\partial}\mass_l}{{\partial}r_{{\lambda}_1}}\right)^{2}} {({\Delta}r_{{\lambda}_1})^2}+{\left(\frac{{\partial}\mass_l}{{\partial}r_{{\lambda}_2}}\right)^{2}} {({\Delta}r_{{\lambda}_2})^2}
\end{equation}
and
\begin{equation}
({\Delta}D_{ol})^2={\left(\frac{{\partial}D_{ol}}{{\partial}r_{{\lambda}_1}}\right)^{2}} {({\Delta}r_{{\lambda}_1})^2}+{\left(\frac{{\partial}D_{ol}}{{\partial}r_{{\lambda}_2}}\right)^{2}} {({\Delta}r_{{\lambda}_2})^2}.
\end{equation}
The transverse speed can be written as $v=R_e/t_0$, so that the error in
transverse speed
becomes
\begin{equation}
({\Delta}v)^2={\left(\frac{dv}{dR_e}\right)^{2}} {({\Delta}R_e)^2} +
{\left(\frac{dv}{dt_0}\right)^{2}} {({\Delta}t_0)^2} \cong {({\Delta}R_e)^2
\over {t_0}^2}
\end{equation}
where $({\Delta}R_e)^2$ can be written explicitly in terms of ${\Delta}\mass_l$
and ${\Delta}D_{ol}$ via Eq.~(\ref{einsteinradius}), and again we neglect
$\Delta{t_0}$, compared to the larger errors.

We simulate the light curve of a microlensing event, in which a $1\Msolar$
main-
sequence star at 8 kpc is lensed by a $0.4\Msolar$ star at 6.25 kpc (i.e., the
near end of the bulge).  We assume
$t_0 = 15$ days and $u_{\rm min} = 0.1$ and that the light curve is sampled
every half hour in ($V$,$I$)
with typical errors of 0.005 mag. Our fitting routine recovers the input
parameters with standard errors of ${\Delta}\mass_l=\pm 0.09\Msolar$,
${\Delta}D_{ol}=\pm 1.5 \mbox{ kpc}$, and ${\Delta}v=\pm 41 \mbox{ km s}^{-1}$
at the
95\% confidence level. For a similar event with $t_0 = 40$ days, the errors
become ${\Delta}\mass_l=\pm 0.05\Msolar$, ${\Delta}D_{ol}=\pm 0.95 \mbox{
kpc}$, and ${\Delta}v=\pm 25 \mbox{ km s}^{-1}$. The errors in the derived lens
properties
scale roughly linearly with the typical photometric errors for the light curve.
It should be noted that since many of the events which exhibit a color shift
are likely to be of relatively long duration, they may also exhibit a
measurable parallax effect due to the Earth's orbital motion (Buchalter \&
Kamionkowski 1995).
In principle, {\em every} microlensing light curve is altered by this effect,
but the small distortion can presently be discerned only for the
longest-duration events (Alcock et al.~1995b). With the errors and sampling
rates discussed above, the parallax effect could be measured for
shorter-duration events and thus provide a further independent constraint on
the lens transverse speed.
These results show that the aforementioned intensive monitoring programs,
designed to discover extra-solar planets, are
ideally suited for measuring the color-shift
effect as well as picking out the parallax effect to far greater accuracy than
current ground-based work. The data from such
surveys can remove entirely the degeneracy among the unknown lens parameters,
not only for the cases where an
orbiting planet is detected, but in a large fraction of events in general.

\section{DISTINGUISHING FROM BLENDED EVENTS}

Some source stars will have unresolved binary companions whose light will be
blended
with that of the source during
the event. The effect of such blending produces a shape distortion entirely
analogous
to a color shift (Griest \& Hu 1992). For reasonable models of the density
distribution and MF
of stars along the line of sight to Baade's Window, it is intuitively evident
that the majority
of blends will be due to faint stars in the bulge. Thus, consider the case
where the lens emits absolutely no light,
but the source star is in a binary.  In this case, the companion
of the source star is unresolved, and light from the companion of the source
star will
cause shape distortions to the light curves similar to those
from the case where the lens emits light.  Therefore, one may
question whether color shifts, if observed, signal that the lens
is a luminous star.  Fortunately, there are several ways of
discriminating between the two scenarios.  The first may be
performed for each individual event.  As discussed in Section 7,
light-curve analysis can potentially yield the
distance and mass of the star---either the lens or the unseen companion
of the source star---contributing the blended light.  If the
distance can be determined to be different from the distance to
the source star, then the blended light is most likely coming
from the lens and not a binary companion to the source. One also
expects that in any given event, the inferred lens mass
will in general be less than the mass of the source. If these
quantities cannot be disentangled,
then one cannot discriminate between the two scenarios, based on the light
curve alone. High resolution imaging of stars along the line of sight, however,
could likely eliminate many
optical superpositions.

There is also a statistical test which can be applied to an
ensemble of color-shifted events.  If the lenses are emitting
the blended light, then we expect the fraction of events of a
given duration which are color shifted to be larger for longer-duration
events.  This is simply because the typical duration of
an event increases for larger lens masses, and the
luminosity of a dwarf increases rapidly with mass.  If the
lenses emit no light and the color shifts are due to blended
light from a companion star, then there should be no correlation
between the event duration and the color shift.  Thus, in this
case, the fraction of events which are color shifted should not
increase as dramatically.  (There will be some increase in the
color-shifted fraction for longer events just because the light
curve in a longer event is better sampled.) This analysis thus
provides a fundamental test of whether the lenses are, in fact, ordinary
stars.

To illustrate, we show in Figure 10 the expected fraction of events of a
given duration whose light curves are distorted due to both of the scenarios
discussed above.  For color shifts from a luminous lens, we consider both bulge
self-lensing (short-dashed curves)
and disk-lensing-bulge events (long-dashed curves),  where all
stars are assumed to lie on the main sequence and follow the logarithmic mass
function discussed in
Section 3.2.  To generate the corresponding dotted curves, we make the same
assumptions about the lens and source mass and spatial
distributions.  However, in this case, we assume that the
lens emits no light and that the color shift is due to
blended light from an unresolved binary companion to the source
star which has a similar mass function.  In the upper panel we assume hourly
sampling and
0.01-mag errors, and in the lower panel we assume 0.5 hour sampling with
0.005-mag errors,
both using ($V$ ,$I$). As this simple example
illustrates, for BB events with $t_0 \geq 10$ days
and DB events with $t_0 \geq 13$ days, the fraction of events which are color
shifted
rises more dramatically with duration if the lenses are
emitting the blended light. Not all source stars are in
binaries, so dot-dashed curves should actually be viewed as an
upper limit to the fraction of events which are color shifted
due to light from an unresolved binary. The Figure also demonstrates our
previous
claim that color-shift analysis is particularly important for large $t_0$.
\begin{figure}
\plotone{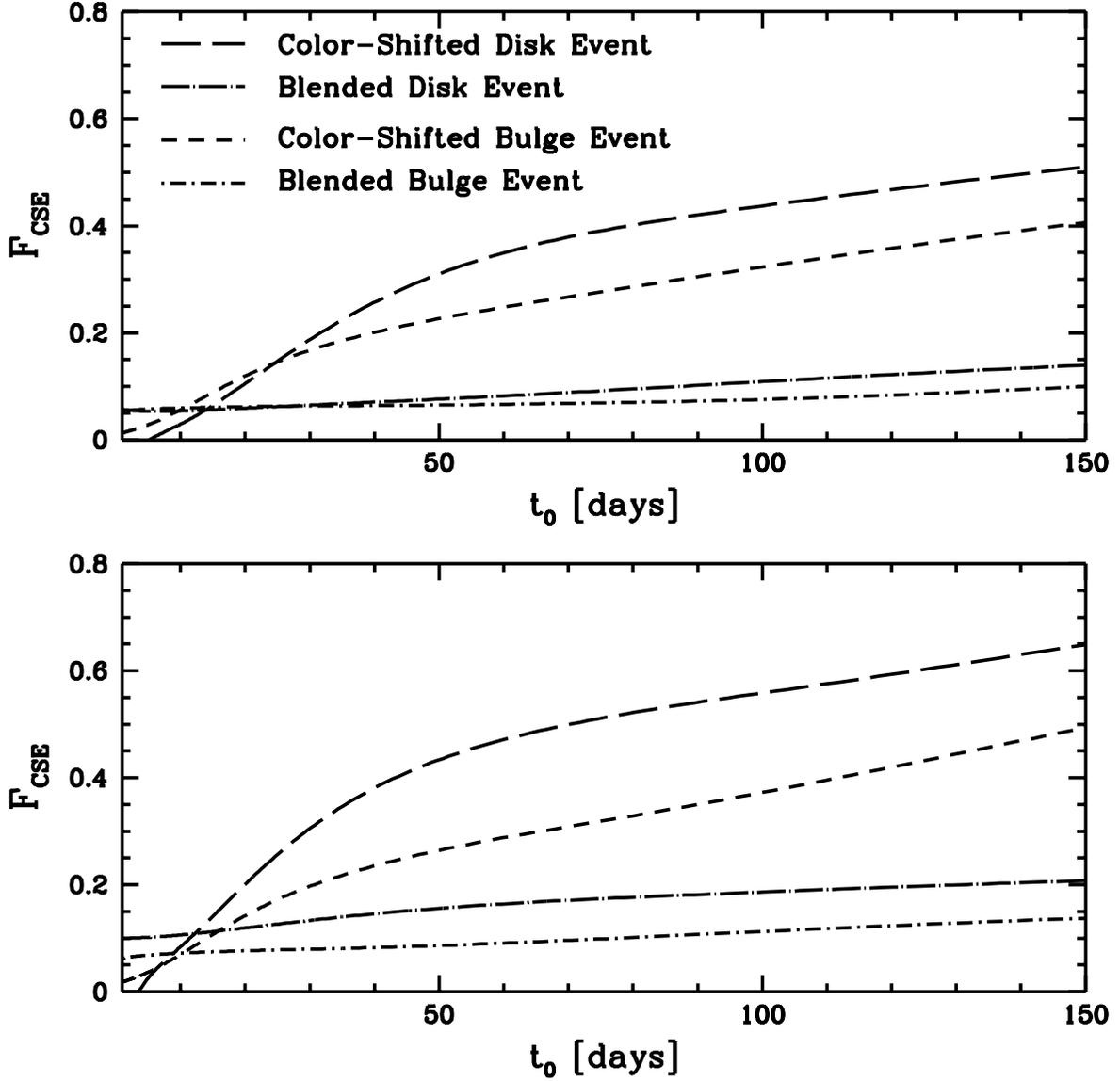}
\caption{Fraction of shape-distorted events seen in ($V$,$I$) as a function of
$t_0$. For the long-dashed and short-dashed
curves, $F_{CSE}$ arises from color shifting by the lens in DD(MS) and DB(MS)
events. The dotted curves
correspond to the same events, but with a dark lens and $F_{CSE}$ arising from
the blended light of a source
companion. The upper panel represents hourly sampling and 0.01-mag errors while
the lower panel corresponds
to 0.5-hour sampling and 0.005-mag errors. The graphs show that optical
blending can be ruled out statistically, and
further illustrate a method to test the assumption that the lenses are stars.}
\end{figure}

\section{CONCLUSION}

The importance of microlensing surveys in the study of Galactic structure has
been firmly established.
The increasing wealth of data on events towards the Galactic bulge makes it
increasingly unlikely
that the lenses are mostly in the form of brown dwarfs and seems to indicate
that they are in
the form of ordinary low-mass stars. If this is the case, they are expected to
give rise to a
characteristic waveband-dependent shape distortion to the light curve, which
can potentially
be distinguished from events where the source is an optical blend. Failing to
account for color shifting can lead to errors in the inferred timescales for
single
events and in the overall duration distribution, particularly for large $t_0$
(Buchalter \& Kamionkowski 1995). Color-shift
analysis can be used with existing data to place upper limits to the lens mass
as a function of distance for
individual events. Our detailed calculation shows
that with frequent and precise observations of events in progress, 5\% to 60\%
of all events arising from various lens-source pairings are expected to exhibit
a color shift, as observed in ($B$,$K$) or ($V$,$I$). These events can be used
to de-convolve the parameters of interest (namely $\mass_l,D_{ol},\mbox{ and
}v$) with reasonable precision. Moreover, it is expected
that many CSEs will also exhibit a measurable parallax shift from such
ground-based observations, which allows
a further independent calculation of the lens characteristics. The
observational requirements for measuring these effects match those of proposed
planetary searches. Such surveys could thus
yield important information for the study of Galactic structure, by
determining important characteristics of the lensing population.

\acknowledgements

We wish to thank H. S. Zhao for providing us with data from his self-consistent
model of
the Galactic bar, and for helpful discussions. We also thank R. Olling for many
insightful suggestions on
constructing a reasonable model for the Galactic disk
as well as numerous helpful comments on a preliminary draft. This work was
supported in part by the U. S. Department of
Energy under contract DE-FG02-92ER40699, and by NASA grants NAGW-2479 and
NAG5-3091.

\end{document}